%% file: hom_pol_rings.tex
\documentclass[useAMS,usenatbib,intlimits,centertags]{mn2e}
\usepackage{graphicx}
\usepackage{amsmath,amssymb}
\usepackage{booktabs}
\usepackage{epsfig}

\usepackage{lscape}
\usepackage{array}
\newcommand{\PBS}[1]{\let\temp=\\#1\let\\=\temp}

\allowdisplaybreaks
\newcommand{\rs}{r_\text{s}}

\renewcommand{\d}{\text{d}}

\begin{document}

 \title{Uniformly Rotating Homogeneous and Polytropic Rings in Newtonian Gravity}

 \author[D.\ Petroff \& S.\ Horatschek]
 {David Petroff\thanks{E-mail: {\tt D.Petroff@tpi.uni-jena.de}} and Stefan Horatschek\thanks{E-mail: {\tt S.Horatschek@tpi.uni-jena.de}}\\
 Theoretisch-Physikalisches Institut, University of Jena, Max-Wien-Platz 1, 07743 Jena, Germany}
 \date{\today}

 \pagerange{\pageref{firstpage}--\pageref{lastpage}} \pubyear{2008}

 \maketitle

 \label{firstpage}

 \begin{abstract}
  An analytical method is presented for treating the problem of a uniformly rotating, self-gravitating
  ring without a central body in Newtonian gravity. The method is based on an expansion about the thin ring limit, where
  the cross-section of the ring tends to a circle. The iterative scheme developed here is applied to homogeneous rings
  up to the 20th order and to polytropes with the index $n=1$ up to the third order. For other
  polytropic indices no analytic solutions are obtainable, but one can apply the method numerically.
  However, it is possible to derive a simple formula relating mass to the integrated pressure
  to leading order without specifying the equation of state.
  Our results are compared with those generated by highly accurate numerical methods to test their accuracy.
 \end{abstract}


 \section{Introduction}
The problem of the self-gravitating ring captured the interest of such renowned scientists as
\citet{Kowalewsky85}, \citet{Poincare85b,Poincare85c,Poincare85d} and \citet{Dyson92, Dyson93}. Each of them tackled
the problem of an axially symmetric, homogeneous ring in equilibrium by expanding it about
the thin ring limit. In particular, Dyson provided a solution to fourth order in the
parameter $\sigma=a/b$, where $a$ provides a measure for the radius of the cross-section
of the ring and $b$ the distance of the cross-section's centre of mass from the axis of rotation.
An important step toward understanding rings with other equations of state was taken by
\citet{Ostriker64,Ostriker64b,Ostriker65}, who studied polytropic rings to first order in $\sigma$
and found a complete solution to this order for an isothermal limit.
\par

First numerical results for homogeneous rings were given by \citet{Wong74},
who was not able to clarify the transition to spheroidal bodies that \citet{Bardeen71}
had supposed would exist.
\citet{ES81} and \citet{EH85} developed improved methods
with which they were able to study the connection
to the Maclaurin spheroids.
Returning to the problem significantly later, \citet*{AKM03} achieved
near-machine accuracy, which allowed them to study bifurcation sequences in detail and correct
erroneous results. It was also possible to extend the problem to non-homogeneous rings and
even to the framework of General Relativity \citep*{Hachisu86,AKM03c,FHA05}.
\par

Through the use of computer algebra, we extend Dyson's basic idea and determine
the solution to the problem of the homogeneous ring up to the order $\sigma^{20}$.
We also present an iterative method for performing a similar expansion about the
thin ring limit for arbitrary equations of state and general results are derived,
confirming and generalizing work that had already been published by \citet{Ostriker64b}.
The application to polytropes is considered and ordinary differential equations (ODEs) are found
that allow for the determination of the mass density. A closed-form solution can only be
found if the value of the polytropic index is $n=1$, and such rings are considered to the order
$\sigma^3$. For other polytropic indices, the ODEs are solved numerically
so that results from the approximate scheme can be compared to highly accurate numerical results
for a variety of equations of state. The numerical solutions considered here are taken from
a multi-domain spectral program, much like the one described in \citet*{AKM03b}, but tailored
to Newtonian bodies with toroidal topologies (see \citealt{AP05} for more information). The solutions
obtained by these numerical methods are extremely accurate and thus provide us with a means
of testing the accuracy of the approximate method.

\section{The Approximation Scheme}\label{approx}

\subsection{The Coordinates}
\begin{figure}
\centerline{\includegraphics{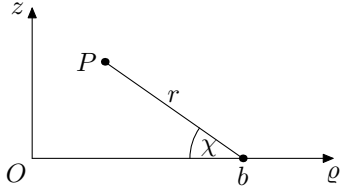}}
\caption{\label{Abb_coord} A sketch providing the meaning of the coordinates $(r,\chi)$ in relation
 to the cylindrical coordinates $(\varrho,z)$.}
\end{figure}

To describe axially symmetric rings, we introduce the polar-like coordinates $(r,\chi,\varphi)$,
which are related to the cylindrical coordinates $(\varrho,z,\varphi)$ by
\begin{align}
\varrho=b-r\cos\chi,\quad z=r\sin\chi,\quad \varphi=\varphi.
\end{align}
For a given value of $\varphi$, constant values of the coordinate $r$ are circles centred about
 $(\varrho=b,z=0)$ and $\chi$ measures the angle
 along any such circle. Fig.~\ref{Abb_coord} provides an illustration of the coordinates.
The surface of the ring will be described by a function $\rs(\chi)$.
In our coordinates, the Laplace operator applied to a function $f=f(r,\chi)$ reads
\begin{align}
\begin{split}
\nabla^2f=&\frac{\partial^2f}{\partial r^2}+\frac1r\frac{\partial f}{\partial r}+\frac1{r^2}\frac{\partial^2f}{\partial\chi^2}\\
          &-(b-r\cos\chi)^{-1}\left(\cos\chi\frac{\partial f}{\partial r}-
            \frac{\sin\chi}r\frac{\partial f}{\partial\chi}\right).
\end{split}
\end{align}
We choose the constant $b$ in such a manner that the centre of mass of the ring's cross-section
coincide with $r=0$, thus implying
\begin{align}\label{com}
 \int_0^{2\pi}\int_0^{\rs(\chi)}\mu r^2\cos\chi\,\d r\,\d\chi=0,
\end{align}
where $\mu$ is the mass density.

\subsection{Basic Equations}
For solving the problem of a self-gravitating fluid in equilibrium, we have to fulfil Laplace's equation
\begin{align}\label{Lap}
 \nabla^2 U_\text{out}=0
\end{align}
outside the fluid and Poisson's equation
\begin{align}\label{Poi}
 \nabla^2 U_\text{in}=4\pi G\mu
\end{align}
inside it, where $U$ is the gravitational potential.
Additionally, we have to satisfy Euler's equation
\begin{align}\label{Eul}
 \mu\frac{\d\bmath v}{\d t}=-\mu\nabla U_\text{in}-\nabla p,
\end{align}
where $\bmath v$ is the velocity of a fluid element and $p$ the pressure.
We consider uniform rotation about the axis $\varrho=0$ with the angular velocity $\mathbf\Omega$
and thus have the velocity field
\begin{align}
 \bmath v=\mathbf\Omega\times\bmath x
\end{align}
leading to
\begin{align}\label{eqi0}
 \nabla\left(U_\text{in}+\int_0^p\frac{\d p'}{\mu(p')}-\frac12\Omega^2\varrho^2\right)=0.
\end{align}
We introduce the pressure function
\begin{align}\label{h}
 h:=\int_0^p\frac{\d p'}{\mu(p')},
\end{align}
which is nothing other than the specific enthalpy in the
case of constant specific entropy. Integration gives
\begin{align}\label{int_Euler}
 U_\text{in}+h-\frac12\Omega^2\varrho^2=V_0,
\end{align}
where $V_0$ is the constant of integration.
At the surface of the ring, where the pressure vanishes, we have
\begin{align}\label{int_Euler_surf}
 U_\text{s}-\left.\frac{1}{2}\Omega^2\varrho^2\right|_\text{s}=V_0.
\end{align}

\subsection{Series Expansions}
The thin ring limit is approached when the ratio
of the inner radius $\varrho_\text{i}$ to the outer one $\varrho_\text{o}$ tends to 1.
In this limit, the cross-section of the ring becomes a circle.
This is the starting point of the approximation.
As an interesting aside, if one considers a ring surrounding a central body, for example a point mass,
the cross-section of the ring can deviate significantly from a circle even if the radius ratio is close to 1,
see Fig.~8 in \citet{AP05}.
\par
We describe the surface of the ring up to order $q$ by the Fourier series
\begin{align}\label{rs}
\rs(\chi)=a\left(1+\sum_{i=1}^q\sum_{k=0}^i\beta_{ik}\cos(k\chi)\sigma^i+o(\sigma^q)\right),
\end{align}
where
\begin{align}
\sigma:=\frac{a}b.
\end{align}
There are no sine terms because of reflectional symmetry with respect to the
equatorial plane, a symmetry necessarily present for fluids in equilibrium
\citep[see][]{Lichtenstein33}. To the leading order, the cross-section
is indeed a circle of radius $a$.
We make a similar ansatz for the mass density, pressure function, pressure and the potential inside the ring
\begin{align}\label{mu_expansion}
  \mu(r,\chi) &= \mu_\text c\left(\sum_{i=0}^q\sum_{k=0}^i \mu_{ik}(y) \cos(k\chi) \sigma^i + o(\sigma^q)\right),\\
  h(r,\chi)   &= \pi G\mu_\text c a^2\left(\sum_{i=0}^q\sum_{k=0}^i h_{ik}(y)\cos(k\chi) \sigma^i+ o(\sigma^q)\right),\label{h_ser}\\
  p(r,\chi)   &= \pi G\mu_\text c^2 a^2\left(\sum_{i=0}^q\sum_{k=0}^i p_{ik}(y)\cos(k\chi) \sigma^i+ o(\sigma^q)\right),\\
  U_\text{in}(r,\chi)&=-\pi G\mu_\text c a^2\left(\sum_{i=0}^q\sum_{k=0}^iU_{ik}(y)\cos(k\chi)\sigma^i+ o(\sigma^q)\right),
\end{align}
where we have introduced the dimensionless radius
\begin{align}
 y := \frac{r}{a}.
\end{align}
The quantity $\mu_\text c$ is chosen to be the mass density
at the point $r=0$ and does not represent the density's maximal value, although it will not differ
significantly from it in general. For the square of the angular velocity, we write
\begin{align}
\Omega^2=\pi G\mu_\text c\left(\sum_{i=0}^{q+2}\Omega_i\sigma^i+o(\sigma^{q+2})\right)
\end{align}
and for $V_0$
\begin{align}
\label{V0_ansatz}
V_0=-\pi G\mu_\text c a^2\left(\sum_{i=0}^q v_i\sigma^i+o(\sigma^q)\right).
\end{align}
\par
Since the potential outside the ring contains logarithmic terms in $r$ it will come as no surprise
that there are $\ln\sigma$ terms in the coefficients of our series. Like Dyson, we
introduce
\begin{align}
\lambda:=\ln\frac{8}\sigma-2.
\end{align}
Because $\lim_{\sigma\to 0}\sigma\lambda^\alpha=0$ for all $\alpha$, this dependence will
not pose a problem for the iteration scheme.

\subsection{Determination of the Coefficients}
 We now present a method for determining $\mu_{qk}$, $\Omega_{q+1}$ and $\beta_{qk}$ given that
 the previous terms in $\sigma^i$ are known.
 \par

 The idea used in Dyson's approximation scheme for homogeneous rings makes use of the Poisson integral
 to determine the gravitational potential in terms of the (still unknown) function $\rs(\chi)$ along
 the axis of rotation \citep{Dyson92}. This is only possible since the mass density is completely determined
 for homogeneous matter
 once the shape of the ring is given. In general, however, it is necessary first to determine $\mu$ to
 the desired order before being able to perform the integral. By taking the divergence of the Euler
 equation \eqref{eqi0}, using \eqref{Poi} and expressing $h$ as a function of $\mu$ using the equation of state,
 we obtain a second order PDE for $\mu$:
  \begin{align}\label{Laplace_Euler}
   4\pi G\mu + \nabla^2 h -2 \Omega^2= 0.
  \end{align}
 Expanding in terms of $\sigma$ and requiring that
 the equations be satisfied for each power in $\sigma$ and each term in the Fourier
 expansion then results in ODEs for $\mu_{qk}(y)$ once an
 equation of state has been specified. These functions must
 be regular at the origin and chosen such that $\mu_{00}(0)=1$ and $\mu_{ik}(0)=0$ for all other
 $i$ and $k$ so as to be consistent with the choice $\mu(0,\chi)=\mu_\text c$. For $k=0$, this
 condition suffices to determine the function uniquely. For $k=1,2,\ldots,q$, the remaining constants
 in the solutions of the ODEs are found by requiring that the pressure and thus pressure function
 vanish at the surface. Demanding this
 for each of the coefficients in a Fourier expansion, provides $q+1$ equations for the remaining $q$ constants.
 The additional equation can be used to determine $\beta_{q0}$.

Although the Poisson integral is valid everywhere, for technical reasons,
we first calculate the potential on the axis of symmetry only.
After this, we determine the potential outside the ring and
in particular along its surface, where \eqref{int_Euler_surf} must hold.
\par
We label the coordinates for a point on the axis
$(R,\chi_R)\equiv(r,\chi)$, from which
\begin{align}\label{axis}
 b=R\cos(\chi_R)
\end{align}
follows. The axis potential is
 \begin{align}
  \begin{split}
  U&_\text{axis}(R)= -2\pi G\int_0^{2\pi}\int_0^{\rs(\chi')}\!\!\!\frac{\mu\,(b-r\cos \chi')r}{\sqrt{R^2+r^2-2Rr\cos\psi}}\,\d r\,\d\chi'\\
     &= -2\pi G\int_0^{2\pi}\!\int_0^{\rs(\chi')}\!\!\!\mu(b-r\cos\chi') \sum_{l=0}^\infty \left(\frac{r}{R}\right)^{l+1}\!\!P_l(\cos \psi)\,
         \d r\,\d\chi' \\
     &=:-2\pi^2 G\mu_\text c a^2\sum_{l=1}^{\infty}\frac{(2l-1)!!}{2l-1}\left(\frac{a}{R}\right)^{2l-1}\frac{A_l}{\sigma^l}\\
     &= -2\pi^2 G\mu_\text c a^2\left(\sum_{l=1}^{q+1}\frac{(2l-1)!!}{2l-1}\left(\frac{a}{R}\right)^{2l-1}\frac{A_l}{\sigma^l}
          + o(\sigma^q)\right),
 \end{split}
 \end{align}
 where $\psi:=\chi'-\chi_R$ and $P_l$ denote the Legendre polynomials.
 Please note that the expansion in terms of powers of $1/R$ indicated in the
 last line is not trivial, since there is an $R$-dependence hidden in the terms
 with $\psi$. Because of reflectional symmetry, there are only terms with odd powers in $1/R$.
 We expand $A_l$ with respect to $\sigma$
 \begin{align}\label{def:alpha}
  A_l=\sum_{i=l-1}^q\alpha_{li}\sigma^i+o(\sigma^q).
 \end{align}
 Using $U_\text{axis}$, we can then find the potential
 anywhere in the vacuum region. To do so, we first introduce a set of axially symmetric
 solutions to Laplace's equation that vanish at infinity. We define
 \begin{align}
  I_1(\varrho,z) &:=\int_0^\pi\!\frac{\d\varphi}{\sqrt{b^2+\varrho^2+z^2-2b\varrho\cos\varphi}},
 \end{align}
 which is nothing other than a multiple of the potential of a circular line
 of mass with radius $b$, centred around the axis. In $(r,\chi,\varphi)$-coordinates it reads
\begin{align}
 I_1(r,\chi)=\frac{2K\!\left(\sqrt{\frac{4b^2-4br\cos\chi}{4b^2-4br\cos\chi+r^2}}\right)}
                  {\sqrt{4b^2-4br\cos\chi+r^2}},
 \end{align}
 where $K$ denotes the complete elliptic integral of the first kind,
 \begin{align}
 K(k):=\int_0^{\frac{\pi}2}\frac{\d\theta }{\sqrt{1-k^2\sin^2\theta}}.
 \end{align}
 Because the difference
 of two such solutions with different $b$'s also satisfies Laplace's equation, it is clear that
\begin{align}
 I_l(r,\chi):=\left(\!-\frac{1}{b}\frac{\d}{\d b}\!\right)^{l-1}\!I_1(r,\chi),
 \end{align}
 where
 \begin{align}
\frac{\d}{\d b}=\frac{\partial}{\partial b}+\cos\chi\frac{\partial}{\partial r}-\frac{\sin\chi}r\frac{\partial}{\partial\chi},
 \end{align}
 is also a solution of Laplace's equation.
 Next we note that along the axis we have
 \begin{align}
  I_l(R) = \frac{\pi(2l-1)!!}{(2l-1)R^{2l-1}}.
 \end{align}
 It then follows that the potential in the vacuum region is
 \begin{align}\label{U_out}
  U_\text{out}(r,\chi)=-2\pi G\mu_\text c a^2\left(\sum_{l=1}^{q+1}a^{2l-1}\sigma^{-l}
  A_lI_l+o(\sigma^q)\right),
 \end{align}
 since this expression satisfies Laplace's equation, vanishes at infinity and
 has the correct value along the axis. For calculating the potential at the body's surface
 we expand $I_l(r,\chi)$ for $r<b$. For example we get
 \begin{align}
  \label{I1}
   I_1(r,\chi)=\frac1b\left[\ln\left(\frac{8b}r\right)+
               \frac{[\ln(8b/r)-1]\cos\chi}{2}\frac{r}b+o\left(\frac{r}b\right)\right].
 \end{align}
 After evaluating these equations for $I_l(r,\chi)$ at the surface
 $r=\rs(\chi)$, we use \eqref{U_out} to find the coefficients $\phi_{ik}$ in the expansion
 \begin{align}
 \label{Us}
 U_\text{s}(\chi)=-2\pi G\mu_\text c a^2\left(\sum_{i=0}^q\sum_{k=0}^i\phi_{ik}\cos(k\chi)\sigma^i
                    +o(\sigma^q)\right).
 \end{align}
 The coefficients $\phi_{qk}$ still depend on $q$ unknown $\beta_{qk}$ (remember that $\beta_{q0}$
 is already known). By comparing the coefficients of $\cos(k\chi)$ ($k=1,2,\ldots,q$),
 equation \eqref{int_Euler_surf} provides $q$ equations. Together with equation \eqref{com} to the relevant
 order in $\sigma$, we can solve for $\Omega_{q+1}$ and the aforementioned $\beta_{qk}$. The absolute
 term ($\cos(0\chi)$) gives a relation between $\Omega_{q+2}$ and $v_q$.

 \section{General Results to First Order}\label{gen_results}

 The approximation scheme described above allows us to draw certain conclusions even without
 specifying the equation of state, thus generalizing results that were published for polytropes
 by \citet{Ostriker64b}. To leading order in $\sigma$, where nothing depends on the
 angle $\chi$, the ring (here a torus) is equivalent to an infinitely long cylinder, a problem
 that was studied by \cite{CF53, Ostriker64}. Equation \eqref{Laplace_Euler} now reads
 \begin{align}\label{h00}
  \left(\frac{\d^2}{\d y^2}+ \frac{1}{y}\frac{\d}{\d y}\right)h_{00} + 4\mu_{00} = 0,
 \end{align}
 since the integrated Euler equation \eqref{int_Euler} tells us that
 \begin{align}\label{Om_order}
  \frac{\Omega^2}{G\mu_\text c} = o(\sigma^2)
 \end{align}
 must hold, i.e.
 \begin{align}\label{Om_01}
  \Omega_0=\Omega_1=0.
 \end{align}
 \par
\begin{figure}
  \centerline{\includegraphics{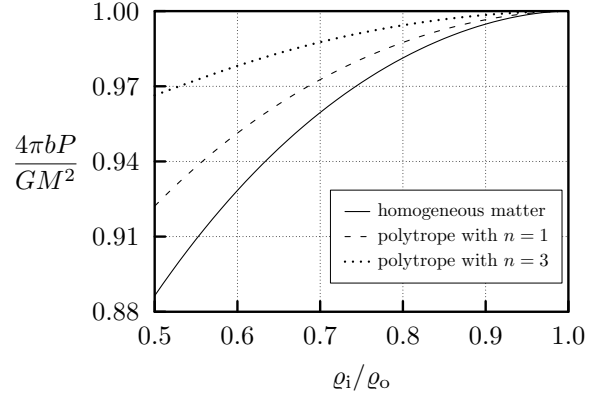}}
  \caption{Numerical examples demonstrating how $4\pi b P/GM^2$ tends to 1 in the thin ring
           limit for various equations of state, cf.\ equation \eqref{eq:P_M2}. \label{P_M2}}
 \end{figure}

 At the surface of the ring $r=\rs$, the pressure vanishes, corresponding
 to $h(r=\rs)=0$, and we thus find
 \begin{align}\label{h00(1)}
  h_{00}(1)=0
 \end{align}
 and
 \begin{align}\label{eq:beta11}
  \beta_{11} = \left.-h_{11}\left(\frac{\d h_{00}}{\d y}\right)^{-1}\right|_{y=1}.
 \end{align}
 By multiplying \eqref{h00} by $\pi^2 \mu_\text c a^2b y$ and integrating from 0 to 1, one finds that the
 mass $M$ to leading order can be related to the derivative of $h_{00}$ at the point $y=1$ according to
 \begin{align}\label{mass_h00}
  M = 4\pi^2 \mu_\text c a^2 b\int_0^1 \mu_{00}y\,\d y = -\pi^2\mu_\text c a^3\sigma^{-1}\left.\frac{\d h_{00}}{\d y}\right|_{y=1}.
 \end{align}
 A particularly interesting relation involving the square of the mass can be derived by considering
 the integral over the pressure
 \begin{align}
   P := 2\pi \int_0^{2\pi}\int_0^{r_\text s(\chi)} pr (b-r\cos\chi)\,\d r \,\d \chi.
 \end{align}
 To leading order, upon taking \eqref{h00} into account, this integral reads
 \begin{align}
  \begin{split}\label{eq:P_M2}
   P &= 4\pi^3 G\mu_\text c^2 a^4 b\int_0^1 p_{00} y\,\d y \\
     &= -2\pi^3 G\mu_\text c^2 a^4 b\int_0^1 \frac{\d p_{00}}{\d y} y^2\,\d y \\
     &= -2\pi^3 G\mu_\text c^2 a^4 b \int_0^1 \mu_{00} \frac{\d h_{00}}{\d y} y^2\,\d y\\
     &= 8\pi^3 G \mu_\text c^2 a^4 b \int_0^1 \mu_{00}y \left(\int_0^y \mu_{00}  y'\,\d y'\right)  \d y\\
     &= 4\pi^3 G \mu_\text c^2 a^4 b \left(\int_0^1 \mu_{00}y\, \d y\right)^{\!2}\\
     &= \frac{GM^2}{4\pi b}.
  \end{split}
 \end{align}
 Numerical examples demonstrating how $4\pi b P/GM^2$ approaches 1 in the thin ring limit
 for various equations of state can be found in Fig.~\ref{P_M2}.
 \par
\begin{figure}
  \centerline{\includegraphics{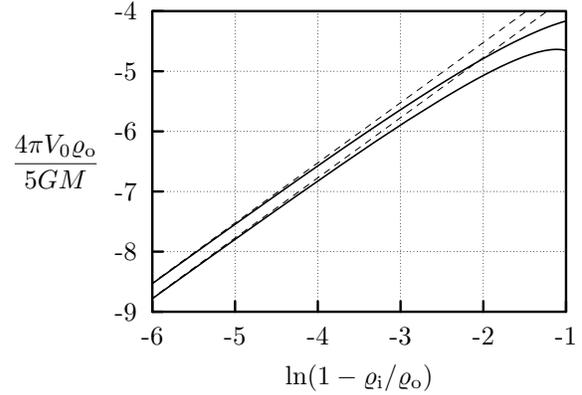}}
  \caption{Numerical ring sequences (solid lines) for homogeneous matter (upper curve)
           and polytropes with $n=5$ (lower curve) are plotted for rings approaching the thin ring limit.
           The asymptotic behaviour as given by equation \eqref{b_V_0_M} is indicated by the dashed lines.\label{n1n5}}
 \end{figure}

 Equation \eqref{com} tells us that
 \begin{align}\label{com_sig1}
  \beta_{11}\mu_{00}(1) + \int_0^1 \mu_{11} y^2 \, \d y =0
 \end{align}
 holds. The terms from the expansions \eqref{U_out} and \eqref{def:alpha} of the potential in the vacuum that we need
 here are
 \begin{align}
  \alpha_{10} &= 2\int_0^1 \mu_{00} y\,\d y = \frac{M\sigma}{2\pi^2 \mu_\text c a^3}\intertext{and}
  \label{def_g}
  \alpha_{21} &= \frac{gM\sigma}{2\pi^2\mu_\text c a^3},\qquad g:=-\frac{\pi^2\mu_\text c a^3}{M\sigma}\int_0^1 \mu_{00} y^3\, \d y,
 \end{align}
 where we used \eqref{com_sig1}.
 The coefficient $\Omega_2$ from the expansion of the square of the angular velocity follows
 from the coefficient in front of $\cos(\chi)\sigma$ in \eqref{int_Euler_surf} and reads
 \begin{align}\label{eq:Om2}
  \Omega_2 &= \alpha_{10}(1+\lambda-2\beta_{11}) + 2\alpha_{21}.
 \end{align}
 According to \eqref{U_out} and \eqref{I1}, the potential at the surface is
 \begin{align}\label{Us_general}
  U_\text s = -\frac{GM}{\pi b}(\lambda+2+ o(1))
 \end{align}
 and the constant of integration $V_0$ then follows immediately from the leading order of
 \eqref{int_Euler_surf}
 \begin{align}\label{V0}
  \begin{split}
   V_0 &= \left.\left(U_\text{out} - \frac{1}{2}\Omega^2\varrho^2\right)\right|_{r=\rs}\\
      &= -\frac{GM}{2\pi b}\left(\frac{5\lambda+9}{2} +g -\beta_{11}+ o(1) \right).
  \end{split}
 \end{align}
 The term $g-\beta_{11}$ appearing in the above equation can be treated further by considering
 the rotational energy $T$ and potential energy $W$ and making use of the virial identity
 \begin{align}\label{VI}
    0 &= 3P+ 2T+W,
 \end{align}
 which then implies
 \begin{align}\label{VI_n1}
      0= 3P+ \frac{GM^2}{\pi b}\left(g-\beta_{11}-\frac{1}{2} \right)
        - 2\pi^3 G\mu_\text c^2 a^4 b\int_0^1 \mu_{00} h_{00}y\, \d y.
 \end{align}
 By restricting ourselves to the polytropic equation of state (see \eqref{poly}), we
 can rewrite the above integral to read
 \begin{align}
  \begin{split}
    \int_0^1 \mu_{00}h_{00}y\,\d y &= \frac{(n+1)K\mu_\text c^{1/n-1}}{\pi Ga^2}\int_0^1 \mu_{00}^{1+1/n} y\, \d y\\
    &=(n+1) \int_0^1 p_{00} y\,\d y
      =\frac{(n+1)M^2}{16\pi^4 \mu_\text c^2 a^4 b^2},
  \end{split}
 \end{align}
 where the last step follows from \eqref{eq:P_M2}. Putting this expression into \eqref{VI_n1}
 and using \eqref{eq:P_M2} again then yields
 \begin{align}\label{g_minus_beta11}
  g-\beta_{11} = \frac{n-1}{8}.
 \end{align}
 Taking into account $1-\varrho_\text i/\varrho_\text o = 2\sigma$, which holds to leading order,
 we can use \eqref{eq:Om2} to write
 \begin{align}\label{b_Om2_M}
  \frac{2\pi b^3 \Omega^2}{GM}+\ln\left(1-\frac{\varrho_\text i}{\varrho_\text o}\right) \to \frac{n-5}{4} + \ln 16
 \end{align}
 and \eqref{V0} can be written as
 \begin{align}\label{b_V_0_M}
  \frac{4\pi b V_0}{5GM}-\ln\left(1-\frac{\varrho_\text i}{\varrho_\text o}\right) \to \frac{5-n}{20} - \ln 16
 \end{align}
 for polytropes in the thin ring limit. Similar equations can be derived for $J$, $T$, $P$ and via the
 virial identity for $W$ \citep[see][]{Ostriker64b}. These equations also hold for homogeneous bodies ($n=0$)
 and numerical examples demonstrating the behaviour \eqref{b_V_0_M} are provided in Fig.~\ref{n1n5}.

 \section{Homogeneous Rings}

 Homogeneous matter is defined by the simple mass distribution
 \begin{align}\label{hom_eos}
  \mu=\mu_\text c,
 \end{align}
 and we can write
 \begin{align}\label{mu_hom}
   \mu_{ik}=\left\{\begin{array}{ll}1&\text{if~}i=k=0\\
                                    0&\text{otherwise}
                  \end{array}\right. .
 \end{align}
 The integral \eqref{h} for the pressure function gives $h=p/\mu_\text c$, which means that
 $h_{ik}=p_{ik}$.
 \par
 For homogeneous rings, it is possible to choose
 \begin{align}
  \beta_{i0}=0
 \end{align}
 without loss of generality.

 \subsection{The Zeroth Order: $\bmath{\sigma^0}$}
 To leading order, the ring is truly a torus
 \begin{align}\label{rs_q0}
  \rs=a[1+o(1)]
 \end{align}
 and equation \eqref{com} is thus automatically fulfilled. The potential along the
 axis is that of a solid
 torus\footnote{$E$ denotes the complete elliptic integral of the second kind,
 $E(k):=\int_0^{\frac{\pi}2}\sqrt{1-k^2\sin^2\theta}\,\d\theta$.}, which can be written down explicitly
 \begin{align}\label{Uaxis_q0}
 \begin{split}
  U_\text{axis}=-\frac{8\pi G\mu_\text c a^3}{3R}\bigg[\left(1+\frac{R^2}{a^2}\right)E\left(\frac{a}{R}\right)\sigma^{-1}\\
  +\left(1-\frac{R^2}{a^2}\right)K\left(\frac{a}{R}\right)\sigma^{-1}+o(\sigma^{-1})\bigg].
 \end{split}
 \end{align}
 An expansion in powers of $1/R$ then gives
 \begin{align}
  A_1=1 + o(1) \quad\Longrightarrow\quad \alpha_{10}=1.
 \end{align}
We can extend this axis potential
to the whole exterior of the ring via \eqref{U_out}, and
expand it on the surface \eqref{rs_q0} by using
\eqref{I1} in $\sigma$. We get
\begin{align}
\label{Us_q0}
U_\text{s}=-2\pi G\mu_\text c a^2\left[\lambda+2 + o(1)\right],
\end{align}
which implies
\begin{align}
\phi_{00}=\lambda+2.
\end{align}
Evaluating equation \eqref{int_Euler_surf} confirms \eqref{Om_01} and provides the relation
\begin{align}\label{Om2_v0}
\Omega_2=2v_0-4(\lambda+2).
\end{align}
This equation cannot be further evaluated until the next order $q=1$.
\par
To calculate the inner structure, we have to solve equation
\eqref{h00}, where the mass density is given by
\eqref{mu_hom}. The solution is
\begin{align}
h_{00}=-y^2+C_1\ln y+C_2.
\end{align}
At the centre, the pressure function has to be regular, thus $C_1=0$, and it must vanish
at the surface \eqref{h00(1)}, thus $C_2=1$.
The integrated Euler equation \eqref{int_Euler} gives us the potential to this order
\begin{align*}
U_{00}=h_{00}+v_0-\frac{\Omega_2}2,
\end{align*}
where both $v_0$ and $\Omega_2$ are unknown, but the required combination of them is given by \eqref{Om2_v0}
and we can conclude
\begin{align*}
U_{00}=2\lambda+5-y^2,
\end{align*}
which indeed gives \eqref{Us_q0} at the surface.

\subsection{The First Order: $\bmath{\sigma^1}$}
The surface function reads
\begin{align}
\rs=a[1+\beta_{11}\sigma\cos\chi+o(\sigma)]
\end{align}
and equation \eqref{com} becomes
\begin{align}
0=\int_0^{2\pi}\rs^3\!\cos\chi\,\d\chi
=a^3[3\pi\beta_{11}\sigma+o(\sigma)].
\end{align}
This gives
\begin{align}
\beta_{11}=0,
\end{align}
cf.\ \eqref{com_sig1},
which means that we have again the surface function \eqref{rs_q0}, but one step further
in $\sigma$:
\begin{align}\label{rs_q1}
\rs=a[1+o(\sigma)].
\end{align}
Accordingly, we get the same potential as \eqref{Uaxis_q0}, but with $o(1)$ instead of $o(\sigma^{-1})$.
The expansion in terms of powers of $1/R$ gives
\begin{align}
A_1=1+o(\sigma)\quad\text{and}\quad A_2=-\frac{\sigma}8+o(\sigma).
\end{align}
The new coefficients are
\begin{align}
\alpha_{11}=0\quad\text{and}\quad\alpha_{21}=-\frac18.
\end{align}
Again we are able to calculate the potential at the body's surface \eqref{rs_q1} via \eqref{U_out},
\eqref{I1} and the corresponding equation for $I_2$.
We get
\begin{align}
\label{Us_q1}
U_\text{s}=-2\pi G\mu_\text c a^2\left[\lambda+2+\left(\frac{\lambda}2+\frac{3}8\right)\sigma\cos\chi+o(\sigma)\right],
\end{align}
which gives us the new coefficients
\begin{align}
\phi_{10}=0\quad\text{and}\quad
\phi_{11}=\frac{\lambda}{2}+\frac{3}8.
\end{align}
Plugging the expansion
\begin{align}
\label{v_squar}
\left.\frac12\Omega^2\varrho^2\right|_\text{s}=\pi G\mu_\text c a^2
  \left[\frac{\Omega_2}2+\left(\frac{\Omega_3}2-\Omega_2\cos\chi\right)\sigma+o(\sigma)\right]
\end{align}
together with \eqref{V0_ansatz} and \eqref{Us_q1} into \eqref{int_Euler_surf} yields
two new equations when one collects the coefficients in $\sigma^i\cos(k\chi)$:
\begin{align}
 \Omega_2&= \lambda+\frac34,\\
 \Omega_3&= 2v_1. \label{Om3_v1}
\end{align}
Using equation \eqref{Om2_v0} then gives
\begin{align}
 v_0 = \frac52\lambda+\frac{35}8.
\end{align}
Equation \eqref{Om3_v1} cannot be further evaluated until the next order $q=2$.
\par
For the mass and the angular momentum, we get
\begin{align}
M&=2\pi^2\mu_\text c a^3[\sigma^{-1}+o(1)]\intertext{and}
J&=\sqrt{\pi^5G\mu_\text c^3\left(4\lambda+3\right)}\,{a}^{5}[\sigma^{-2}+o(\sigma^{-2})].
\end{align}
To leading order, $\varrho_\text{i}/\varrho_\text{o}=1-2\sigma$ holds,
and we can conclude that
\begin{align}
\lim_{\varrho_\text{i}/\varrho_\text{o}\to 1}\left[\frac{4\pi bV_0}{5GM}-\ln\left(1-\frac{\varrho_\text{i}}{\varrho_\text{o}}\right)\right]=\frac14-\ln 16,
\end{align}
as we already saw in \eqref{b_V_0_M}, see also equation (11) in \citet*{FHA05}.
\par
To calculate the inner quantities, we have to find a solution to \eqref{Laplace_Euler}.
The ansatz \eqref{h_ser} leads to the ODEs
\begin{align}
&\frac{\d^2h_{10}}{\d y^2}+\frac1y\frac{\d h_{10}}{\d y}=0\intertext{and}
&\frac{\d^2h_{11}}{\d y^2}+\frac1y\frac{\d h_{11}}{\d y}-\frac{h_{11}}{y^2}+2y=0.
\end{align}
Note that the $2y$ term in the second equation results from $h_{00}=1-y^2$, which is already known.
The solutions of these equations that are regular at the centre and obey $h(\rs)=0$ are
\begin{align}
h_{10}&=0\intertext{and}
h_{11}&=\frac{y}{4}(1-y^2).
\end{align}
Using these results, equation \eqref{int_Euler} then gives
\begin{align}
U_{10}&=0\intertext{and}
U_{11}&=(\lambda+1)y - \frac{y^3}{4}.
\end{align}
\par
After finding the inner potential and pressure, we can calculate the potential energy
\begin{align*}
W=\frac{\mu_\text c}2\int U\,\d V
=-\pi^3G\mu_\text c^2a^5\left[\left(2\lambda+\frac{9}{2}\right)\sigma^{-1}+o(1)\right],
\end{align*}
the rotational energy
\begin{align*}
T=\frac{\mu_\text c\Omega^2}2\int\varrho^2\,\d V
=\pi^3G\mu_\text c^2a^5\left[\left(\lambda+\frac34\right)\sigma^{-1}+o(\sigma^{-1})\right]
\end{align*}
and the integral over the pressure
\begin{align}
P=\int p\,\d V
=\pi^3G\mu_\text c^2a^5\left(\sigma^{-1}+o(1)\right)
\end{align}
to first order.
We see that the virial theorem \eqref{VI} is fulfilled at the leading order.
 \par
For further results see Tables~\ref{tab_omega}--\ref{tab_alpha}.

\begin{table*}
  \centering
  \begin{minipage}{140mm}
  \caption{For a given radius ratio $\varrho_\text i/\varrho_\text o=0.9$,
physical quantities to different orders in $q$ and numerically determined values
           are compared to the
           values for $q=20$:
$\hat M_{20}=4.6299179884304816293\times 10^{-2}$,
$\hat\Omega^2_{20}=3.2474683264953211610\times 10^{-2}$,
$\hat J_{20}=7.5456215256289320669\times 10^{-3}$,
$\hat P_{20}=1.7862946528142761708\times 10^{-4}$,
$\hat T_{20}=6.7988816964653749490\times 10^{-4}$,
$\hat W_{20}=-1.8956647351373578410\times 10^{-3}$.
  \label{tab09}}
  \begin{tabular}{cccccccc}\toprule
$q$ & $\sigma$ &  $\hat{M}_q/\hat{M}_{20}-1$ & $\hat{\Omega}^2_q/\hat{\Omega}^2_{20}-1$ &
$\hat{J}_q/\hat{J}_{20}-1$ & $\hat{P}_q/\hat{P}_{20}-1$ & $\hat{T}_q/\hat{T}_{20}-1$ & $\hat{W}_q/\hat{W}_{20}-1$\\ \midrule
$ 1$ & $0.053$ & $\phantom{+}1\times 10^{-2}$ & $\phantom{+}1\times 10^{-2}$ & $\phantom{+}2\times 10^{-2}$ & $\phantom{+}3\times 10^{-2}$ & $\phantom{+}2\times 10^{-2}$ & $\phantom{+}2\times 10^{-2}$ \\
$ 2$ & $0.052$ & $\phantom{+}2\times 10^{-4}$ & $\phantom{+}6\times 10^{-4}$ & $\phantom{+}4\times 10^{-4}$ & $\phantom{+}6\times 10^{-3}$ & $-1\times 10^{-3}$ & $\phantom{+}6\times 10^{-4}$ \\
$ 3$ & $0.052$ & $\phantom{+}2\times 10^{-4}$ & $\phantom{+}1\times 10^{-4}$ & $\phantom{+}2\times 10^{-4}$ & $\phantom{+}4\times 10^{-4}$ & $\phantom{+}3\times 10^{-4}$ & $\phantom{+}3\times 10^{-4}$ \\
$ 4$ & $0.052$ & $\phantom{+}3\times 10^{-6}$ & $-1\times 10^{-5}$ & $-3\times 10^{-6}$ & $\phantom{+}5\times 10^{-5}$ & $-5\times 10^{-5}$ & $-2\times 10^{-5}$ \\
$10$ & $0.052$ & $\phantom{+}6\times 10^{-11}$ & $-9\times 10^{-11}$ & $\phantom{+}4\times 10^{-12}$ & $\phantom{+}5\times 10^{-10}$ & $-6\times 10^{-10}$ & $-3\times 10^{-10}$ \\
$19$ & $0.052$ & $\phantom{+}4\times 10^{-17}$ & $\phantom{+}4\times 10^{-17}$ & $\phantom{+}6\times 10^{-17}$ & $\phantom{+}1\times 10^{-16}$ & $\phantom{+}9\times 10^{-17}$ & $\phantom{+}9\times 10^{-17}$ \\
\midrule[0.2pt]
num& --- & $-4\times 10^{-14}$& $\phantom{+}1\times 10^{-16}$& $-4\times 10^{-14}$& $-7\times 10^{-14}$& $-4\times 10^{-14}$& $-3\times 10^{-14}$\\
\bottomrule
  \end{tabular}
  \end{minipage}
 \end{table*}

\begin{table*}
  \centering
  \begin{minipage}{140mm}
  \caption{For a given radius ratio $\varrho_\text i/\varrho_\text o=0.5$,
physical quantities to different orders in $q$
are compared to the numerically determined values
$\hat M_\text{num}=0.7201292$,
$\hat\Omega^2_\text{num}=0.5467604$,
$\hat J_\text{num}=0.3247949$,
$\hat P_\text{num}=0.04874713$,
$\hat T_\text{num}=0.1200820$,
$\hat W_\text{num}=-0.3864053$.
  \label{tab05}}
  \begin{tabular}{cccccccc}\toprule
$q$ & $\sigma$ & $\hat{M}_q/\hat{M}_\text{num}-1$ & $\hat{\Omega}^2_q/\hat{\Omega}^2_\text{num}-1$ &
$\hat{J}_q/\hat{J}_\text{num}-1$ & $\hat{P}_q/\hat{P}_\text{num}-1$ & $\hat{T}_q/\hat{T}_\text{num}-1$ & $\hat{W}_q/\hat{W}_\text{num}-1$\\ \midrule
$ 1$ & $0.33$ & $2.8\times 10^{-1}$ & $2.3\times 10^{-1}$ & $4.2\times 10^{-1}$ & $8.6\times 10^{-1}$ & $4.6\times 10^{-1}$ & $6.1\times 10^{-1}$ \\
$ 2$ & $0.30$ & $5.1\times 10^{-2}$ & $5.9\times 10^{-2}$ & $6.5\times 10^{-2}$ & $2.5\times 10^{-1}$ & $2.6\times 10^{-2}$ & $1.1\times 10^{-1}$ \\
$ 3$ & $0.30$ & $5.1\times 10^{-2}$ & $4.2\times 10^{-2}$ & $5.8\times 10^{-2}$ & $1.3\times 10^{-1}$ & $8.1\times 10^{-2}$ & $1.0\times 10^{-1}$ \\
$ 4$ & $0.30$ & $1.7\times 10^{-2}$ & $1.3\times 10^{-2}$ & $1.9\times 10^{-2}$ & $5.4\times 10^{-2}$ & $8.4\times 10^{-3}$ & $2.6\times 10^{-2}$ \\
$10$ & $0.29$ & $1.2\times 10^{-3}$ & $1.0\times 10^{-3}$ & $1.3\times 10^{-3}$ & $3.2\times 10^{-3}$ & $1.0\times 10^{-3}$ & $1.9\times 10^{-3}$ \\
$20$ & $0.29$ & $2.6\times 10^{-5}$ & $2.5\times 10^{-5}$ & $3.0\times 10^{-5}$ & $7.2\times 10^{-5}$ & $2.7\times 10^{-5}$ & $4.4\times 10^{-5}$ \\
\bottomrule
  \end{tabular}
  \end{minipage}
 \end{table*}

\begin{table*}
  \centering
  \begin{minipage}{140mm}
  \caption{For a given radius ratio $\varrho_\text i/\varrho_\text o=0.2$,
physical quantities to different orders in $q$
are compared to the numerically determined values
$\hat M_\text{num}=0.9424$,
$\hat\Omega^2_\text{num}=0.9844$,
$\hat J_\text{num}=0.4545$,
$\hat P_\text{num}=0.07865$,
$\hat T_\text{num}=0.2255$,
$\hat W_\text{num}=-0.6869$.
  \label{tab02}}
  \begin{tabular}{cccccccc}\toprule
$q$ & $\sigma$ & $\hat{M}_q/\hat{M}_\text{num}-1$ & $\hat{\Omega}^2_q/\hat{\Omega}^2_\text{num}-1$ &
$\hat{J}_q/\hat{J}_\text{num}-1$ & $\hat{P}_q/\hat{P}_\text{num}-1$ & $\hat{T}_q/\hat{T}_\text{num}-1$ & $\hat{W}_q/\hat{W}_\text{num}-1$\\ \midrule
$ 1$ & $0.67$ & $1.0$ & $7.5\times 10^{-1}$ & $1.6$ & $5.1$ & $1.6$ & $2.8$ \\
$ 2$ & $0.54$& $3.2\times 10^{-1}$ & $3.5\times 10^{-1}$ & $3.9\times 10^{-1}$ & $1.6$ & $3.2\times 10^{-1}$ & $7.6\times 10^{-1}$ \\
$ 3$ & $0.54$& $3.2\times 10^{-1}$ & $2.7\times 10^{-1}$ & $3.5\times 10^{-1}$ & $1.1$ & $5.4\times 10^{-1}$ & $7.2\times 10^{-1}$ \\
$ 4$ & $0.51$ & $1.9\times 10^{-1}$ & $1.7\times 10^{-1}$ & $2.1\times 10^{-1}$ & $6.3\times 10^{-1}$ & $2.1\times 10^{-1}$ & $3.6\times 10^{-1}$ \\
$10$ & $0.47$ & $6.4\times 10^{-2}$ & $6.9\times 10^{-2}$ & $7.1\times 10^{-2}$ & $1.9\times 10^{-1}$ & $8.7\times 10^{-2}$ & $1.2\times 10^{-1}$ \\
$20$ & $0.46$ & $2.2\times 10^{-2}$ & $2.7\times 10^{-2}$ & $2.3\times 10^{-2}$ & $6.0\times 10^{-2}$ & $3.2\times 10^{-2}$ & $4.2\times 10^{-2}$ \\
\bottomrule
  \end{tabular}
  \end{minipage}
 \end{table*}

\subsection{Discussion for Homogeneous Rings}
With this approximation method, we are able to calculate e.g.\ the shape, angular velocity
and pressure of the ring up to arbitrary order in $\sigma$. We have done so up to
the 20th order.
\par
An important question is how good this method is.
In Tables~\ref{tab09}, \ref{tab05} and \ref{tab02} one can see how the dimensionless
quantities
\begin{align}
 \begin{split}
 \frac{\hat{M}}{M}=\frac{1}{\mu_\text c\varrho_\text{o}^3}, \qquad
 \frac{\hat{\Omega}^2}{\Omega^2}&=\frac{1}{G\mu_\text c},     \qquad
 \frac{\hat{J}}{J}=\frac{1}{G^{1/2}\mu_\text c^{3/2}\varrho_\text{o}^5},\\
 \frac{\hat{P}}{P}=\frac{\hat{T}}{T}&=\frac{\hat{W}}{W}=
         \frac{1}{G\mu_\text c^2\varrho_\text{o}^5}
 \end{split}
\end{align}
improve in accuracy with increasing order for different radius ratios.
Especially for thin rings, we get very accurate results. In fact, for rings with radius ratios
$\varrho_\text{i}/\varrho_\text{o}\approx 0.85$ we achieve a precision
which is comparable with that given by the numerical method described in \citet*{AKM03b}.
For larger radius ratios, the accuracy is thus better.
As a co-product, our work provides an independent test of the
accuracy of the numerical method (better than $10^{-13}$ cf.\ Table~\ref{tab09}).
\par
\begin{figure}
\centerline{\epsfig{file=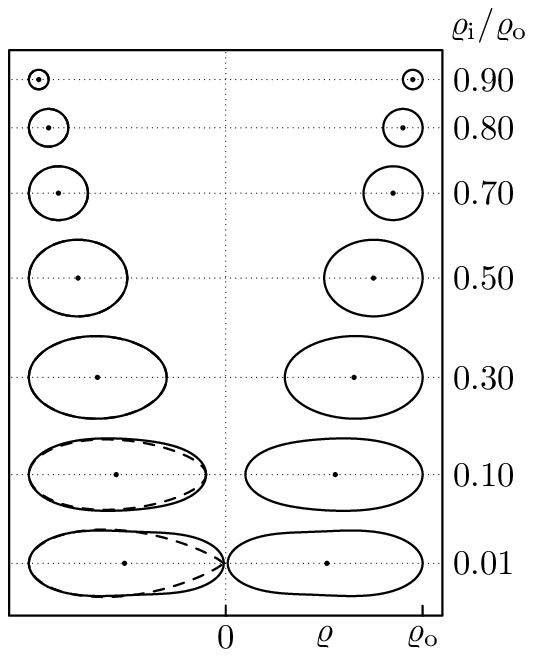}}
\caption{\label{Abb_cs} Meridional cross-sections of homogeneous rings to the order $q=20$ for different radius ratios $\varrho_\text{i}/\varrho_\text{o}$. The $\varrho$- and $z$-axis are scaled identically in such a manner that $\varrho_\text{o}$ has the same value for all the rings. The dot in each ring marks the centre
of mass of the cross-section ($\varrho=b, z=0$ i.e. $r=0$) and the dashed line shows the numerical
result and is indistinguishable from the $q=20$ curve for
$\varrho_\text{i}/\varrho_\text{o}\ga 0.3$.}
\end{figure}
\begin{figure}
\centerline{\epsfig{file=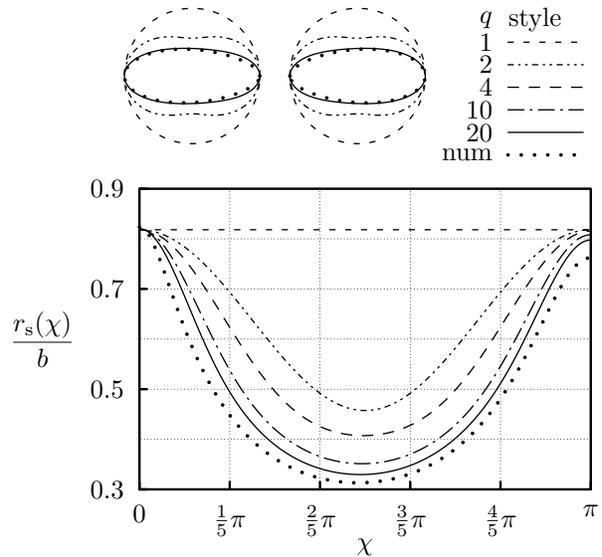}}
\caption{\label{Abb_cs2}
The meridional cross-section and the (dimensionless) surface function $\rs(\chi)/b$
of the homogeneous ring with radius ratio $\varrho_\text{i}/\varrho_\text{o}=0.1$ for different orders $q$ compared
to the numerical result. The surfaces are scaled such that $\varrho_\text{o}$ (and therefore also
$\varrho_\text{i}$) has the same value to all orders.}
\end{figure}
\begin{figure}
\centerline{\epsfig{file=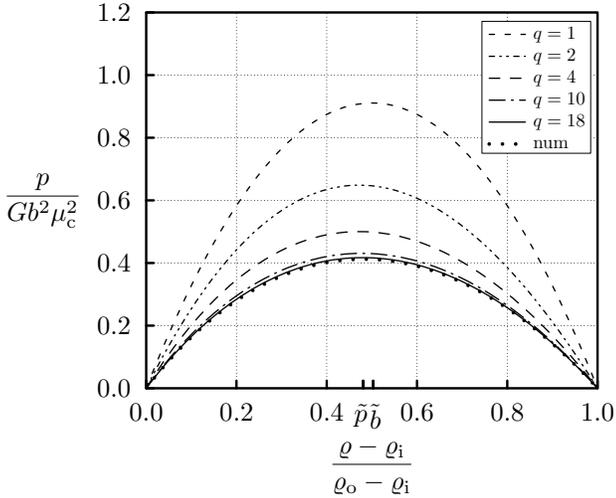}}
\caption{\label{Abb_p} The pressure in the equatorial plane for a homogeneous ring with radius ratio
 $\varrho_\text{i}/\varrho_\text{o}=0.3$ for different orders $q$ compared to the numerical result.
 It is interesting to note that the centre of mass of the cross-section
 does not coincide with the point of maximum pressure.
 To the order $q=18$ we get $\tilde b:=(b-\varrho_\text{i})/(\varrho_\text{o}-\varrho_\text{i})=0.503$
 and $\tilde p:=(\varrho_\text{p,max}-\varrho_\text{i})/(\varrho_\text{o}-\varrho_\text{i})=0.480$, which
 differ by less than 1\% from the numerical values.}
\end{figure}
\begin{figure}
\centerline{\epsfig{file=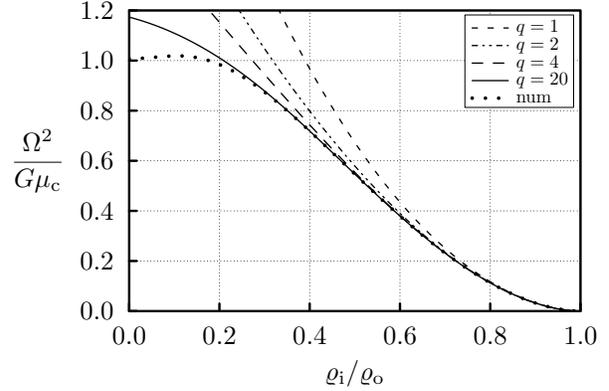}}
\caption{\label{Abb_om} The (dimensionless) squared angular velocity $\Omega^2/G\mu_\text c$ as function of the
 radius ratio $\varrho_\text{i}/\varrho_\text{o}$ for different orders $q$ compared to the numerical
 result for homogeneous rings.}
\end{figure}

The shape of the ring in meridional cross-section for various radius ratios can be found in
Fig.~\ref{Abb_cs}. The curves to order $q=20$ can barely be distinguished from the numerical
ones for $\varrho_\text i/\varrho_\text o\ga 0.3$. As one approaches the transition to spheroidal
topologies ($\varrho_\text i/\varrho_\text o\to 0$), the true curve becomes pointy at the inner edge and is no longer well represented by our Fourier series.
Nevertheless, the shape of the ring is quite well approximated even for $\varrho_\text{i}/\varrho_\text{o}=0.1$, as seen in Fig.~\ref{Abb_cs2}.
The surface function $\rs(\chi)$, which is a constant to leading order, clearly
approaches the numerical one with increasing $q$.
The pressure in the equatorial plane can also be seen to approach the numerically
determined one for $\varrho_\text{i}/\varrho_\text{o}=0.3$ in Fig.~\ref{Abb_p}.
It is interesting to note that the centre of mass does not coincide with the
point of maximum pressure.
In Fig.~\ref{Abb_om} one can get an impression of the accuracy of the approximation over the whole
range of radius ratios and for various values of $q$.
Despite the claims found in \citet{Wong74} that Dyson's perturbative method
diverges for $\sigma>1/3$ \citep[see also the comments in][]{Dyson92,Bardeen71},
these results indicate the opposite.

 \section{Polytropic Rings}
 \subsection{Polytropes with Arbitrary Index $\bmath{n}$}

 The polytropic equation of state is
 \begin{equation}\label{poly}
  p = K\mu^{1+1/n}.
 \end{equation}
 For large/small polytropic indices $n$, the equation is referred to as `soft'/`stiff' and it
 includes homogeneous matter as the limit $\lim_{n \to 0} K^n = 1/\mu = 1/\mu_\text c$.
 From now on, we shall use the terms `homogeneous matter' and `$n=0$' interchangeably.
 For polytropes, \eqref{Laplace_Euler} becomes
 \begin{align}\label{Laplace_Euler_polytrope}
  4\pi G\mu + K(n+1)\nabla^2\left(\mu^{1/n}\right) -2 \Omega^2= 0.
 \end{align}
 Instead of our coordinate $y$, we are now going to make use of a new dimensionless radial coordinate,
 applicable to polytropes
 \begin{align}
  \bar{r} := \frac{G^\frac{1}{2}\mu_\text c^\frac{n-1}{2n}}{K^\frac{1}{2}}  r.
 \end{align}
 To lowest order in $\sigma$, and upon introducing
 \begin{align}
  \tilde\mu:=\mu^{1/n}
 \end{align}
 and the expansion
 \begin{align}
  \tilde\mu   &= \mu_\text c^{1/n}\left(\sum_{i=0}^q\sum_{k=0}^i \tilde\mu_{ik}(\bar{r})\cos(k\chi) \sigma^i + o(\sigma^q)\right),
 \end{align}
 equation \eqref{Laplace_Euler_polytrope} reads, cf.\ \eqref{h00},
 \begin{equation}\label{Lane-Emden}
  \left(\frac{\d^2}{\d \bar{r}^2}+ \frac{1}{\bar{r}}\frac{\d}{\d \bar{r}}\right)\tilde\mu_{00} 
 + \frac{4\pi}{n+1}\tilde\mu_{00}^{ n}= 0.
 \end{equation}
 This equation is sometimes referred to as one of the {\sl generalized Lane-Emden equations (of the first kind)} and
 solutions to it have been derived and studied in e.g.\ \citet{GH00}. No solutions other than for $n=0$ and $n=1$ have been
 found for our particular parameters in closed-form and a discussion using symmetry transformations suggests that they do not
 exist, \citep{Goenner01}. In the limiting case $n=0$, this equation provides an alternative, but less transparent,
 method to the one presented above for treating homogeneous matter to lowest order. We concentrate in the next section on the
 special case $n=1$.

 For other polytropic indices, the iterative method presented here can be
 applied with the help of numerics. By describing the unknown density terms $\tilde\mu_{ik}$ using
 Chebyshev polynomials and expanding all the quantities involved in terms of $\lambda$, equations
 can be formulated for purely numerical coefficients. The equations of the approximation scheme described in
 Section~\ref{approx} must be fulfilled, whereby the ODEs for $\tilde\mu_{ik}$
 are evaluated at collocation points of the Chebyshev polynomials. In general, the density functions
 $\tilde\mu_{00}$ are not analytic at $\bar{r}=\bar a$, meaning that high order polynomials may be necessary to find a
 good approximation of the function desired. We none the less chose this method, since the equations
 involve integrals over the density for which one end-point of integration contains the unknown surface
 function $\rs(\chi)$, making their polynomial representation particularly useful.

 If one is only interested in determining $\bar a$, $\beta_{11}$ and $\Omega_2$, then it is not
 necessary to combine such numerical and algebraic techniques and one can choose
 any numerical method for solving the ODEs. One begins by solving equation \eqref{Lane-Emden}
 numerically for the desired polytropic index $n$, prescribing the `initial conditions'
 $\tilde\mu_{00}(0)=1$ and $\left.\frac{\d}{\d \bar{r}}\tilde\mu_{00}\right|_{\bar{r}=0}=0$.
 For spherical polytropic fluids, a surface of vanishing pressure is known to
 exist only for $n<5$, where the surface for $n=5$ extends out to infinity.
 The situation for polytropic rings is quite different -- it seems that arbitrarily large polytropic
 indices are possible! Numerical solutions to \eqref{Lane-Emden} indicate that the
 density function $\mu_{00}$ indeed falls to zero for large $n$.
 The value of $\bar{r}$ at the first
 zero of the solution is $\bar a$. One then proceeds to solve equation
 \begin{align}
  \left(\frac{\d^2}{\d \bar{r}^2} + \frac{1}{\bar{r}}\frac{\d}{\d \bar{r}} -\frac{1}{\bar{r}^2}\right)\tilde\mu_{11}
   + \frac{4\pi n}{n+1}\tilde\mu_{00}^{n-1}\tilde\mu_{11}=\frac{1}{\bar a}\frac{\d\tilde\mu_{00}}{\d \bar{r}}
 \end{align}
 for $\tilde\mu_{11}$ with the condition $\tilde\mu_{11}(0)=0$ and where $\left.\frac{\d}{\d \bar{r}}\tilde\mu_{11}\right|_{\bar{r}=0}$
 has to be chosen so as to fulfil the centre of mass condition $\eqref{com}$ to first order, which reads
 (cf.\ \eqref{com_sig1})
 \begin{align}
  \begin{split}
    0 &= \int_0^{\bar a} \mu_{11} \bar{r}^2\, \d \bar{r} =  n\int_0^{\bar a} \tilde\mu_{00}^{n-1}\tilde\mu_{11} \bar{r}^2\, \d \bar{r}\\
      \Longrightarrow 0&= \bar a^2\left.\frac{\d\tilde\mu_{11}}{\d \bar{r}}\right|_{\bar{r}=\bar a} - \bar a\tilde\mu_{11}(\bar a)
       + \frac{2}{\bar a}\int_0^{\bar a} \tilde\mu_{00} \bar{r}\, \d \bar{r}.
  \end{split}
 \end{align}
 The constant $\beta_{11}$ can then be found using equation \eqref{eq:beta11}, which now reads
 \begin{align}
  \beta_{11} & = -\frac{\tilde\mu_{11}(\bar a)}{\bar a}\left.\left(\frac{\d\tilde\mu_{00}}{\d \bar{r}} \right)^{-1}\right|_{\bar{r}=\bar a},
 \end{align}
 and $\Omega_2$ is taken from \eqref{b_Om2_M}.
 The behaviour of these coefficients as they depend on the polytropic index $n$ can be found in Table~\ref{beta11_Om2}.
 The table suggests that $\bar a\to \infty$ and $\Omega_2\to 0$ exponentially in $n$ for $n\to\infty$, which is indeed
 known to hold \citep{Ostriker64,Ostriker64b}. The
 behaviour of the specific kinetic energy of a particle in the ring, proportional to $\bar a^2 \Omega_2$ to
 leading order, will be discussed in Section \ref{inf_pol} together with the behaviour of $\beta_{11}$
 for large $n$.
 \begin{table}
  \centering
  \caption{The values of expansion coefficients for the surface function $\rs(\chi)$ and for the squared angular
           velocity are provided up to first order for different polytropic indices $n$. The value of
           $\bar a$ for $n=0$ can be found by solving \eqref{Lane-Emden} with $n=0$ and the conditions
           $\frac{\d}{\d \bar{r}}\tilde\mu_{00}|_{\bar{r}=0}=0$ and $\tilde\mu_{00}(0)=1$ and then locating
           the first zero of $\tilde\mu_{00}$.\label{beta11_Om2}}
  \begin{tabular}{cccc}\toprule
    $n$ & $\bar{a}$                    & $\beta_{11}$ & $\Omega_2$\\ \midrule
     0  & $1/\sqrt{\pi}\approx 0.5642$ & 0            & $\lambda+3/4$ \\
    0.5 & $0.7566$                     & $-0.03537$    & $0.6371\lambda+0.5575$\\
     1  & $0.9594$                     & $-0.07708$    & $0.4318(\lambda+1)$\\
     2  & $1.427$                      & $-0.1731$     & $0.2169\lambda+0.2711$\\
     5  & $3.750$                      & $-0.5118$     & $0.03614\lambda+0.07228$\\
     10 & $15.18$                      & $-1.126$      & $(2.401\lambda+7.804)\times 10^{-3}$\\
     20 & $207.6$                      & $-2.375$      & $(1.362\lambda+7.829)\times 10^{-5}$\\
     30 & $2.661\times 10^3$           & $-3.625$      & $(8.487\lambda+70.02)\times 10^{-8}$\\
     40 & $3.337\times 10^4$           & $-4.875$      & $(5.468\lambda+58.78)\times 10^{-10}$\\
     50 & $4.142\times 10^5$           & $-6.125$      & $(3.577\lambda+47.40)\times 10^{-12}$\\
   \bottomrule
  \end{tabular}
 \end{table}

 We provide a comparison of precise numerical values for various (dimensionless) physical quantities
 with their first and third order equivalents in Table~\ref{various_npoly}.
 The dimensionless quantities, valid for any polytropic index $n>0$ are
 \begin{align}
  \begin{split}
   \frac{\bar{M}}{M}= \frac{G^\frac{3}{2}\mu_\text c^{\frac{n-3}{2n}}}{K^\frac{3}{2}},
       \quad
   \frac{\bar{J}}{J}= \frac{G^2\mu_\text c^\frac{2n-5}{2n}}{K^\frac{5}{2}},
      \quad \frac{\bar{\Omega}}{\Omega}= \frac{1}{G^\frac{1}{2}\mu_\text c^\frac{1}{2}},\\
   \frac{\bar{P}}{P}= \frac{\bar{T}}{T}= \frac{\bar{W}}{W}= \frac{G^\frac{3}{2}\mu_\text c^\frac{n-5}{2n}}{K^\frac{5}{2}},
       \quad
   \frac{\bar{b}}{b} = \frac{\bar{\varrho}}{\varrho}=\frac{G^\frac{1}{2}\mu_\text c^\frac{n-1}{2n}}{K^\frac{1}{2}}.
  \end{split}
 \end{align}
 One can see that the accuracy of the
 method does not depend strongly on the polytropic index and that relative errors at third order 
 are well below one percent for rings with a radius ratio of 0.9.
 \begin{table*}
  \centering
  \begin{minipage}{140mm}
  \caption{Physical quantities to the first and third order in $\sigma$ are
           compared to the correct, numerically determined values for
           given polytropic index $n$ and given radius ratio $\varrho_\text i/\varrho_\text o=0.9$.
           The polytropic indices $n=1.5$ and $n=3$ correspond to a non-relativistic and ultra-relativistic
           completely degenerate Fermi gas respectively.\label{various_npoly}}
  \begin{tabular}{cccccccc}\toprule
    $n$ & $q$ & $\bar{M}$ & $\bar{\Omega}^2$ & $\bar{J}$ & $\bar{P}$ & $\bar{T}$
        & $\bar{W}$\\ \midrule
   1    & 1   & 143.0 & $1.512\times10^{-2}$ & $5.839\times 10^3$ & 89.24 & 359.1 & $-985.8$\\
   1    & 3   & 144.3 & $1.499\times10^{-2}$ & $5.952\times 10^3$ & 90.06 & 364.4 & $-998.9$\\
   1    & num & 144.3 & $1.499\times10^{-2}$ & $5.953\times 10^3$ & 90.07 & 364.5 & $-999.2$\\[1ex]
   1.5  & 1   & 186.9 & $1.094\times10^{-2}$ & $9.836\times 10^3$ & 124.0 & 514.3 & $-1401$\\
   1.5  & 3   & 188.5 & $1.084\times10^{-2}$ & $1.001\times 10^4$ & 125.0 & 521.0 & $-1417$\\
   1.5  & num & 188.6 & $1.084\times10^{-2}$ & $1.001\times 10^4$ & 125.0 & 521.1 & $-1417$\\[1ex]
   3    & 1   & 356.0 & $4.569\times10^{-3}$ & $3.527\times 10^4$ & 263.5 & 1192  & $-3174$ \\
   3    & 3   & 358.7 & $4.524\times10^{-3}$ & $3.581\times 10^4$ & 265.4 & 1204  & $-3205$ \\
   3    & num & 358.8 & $4.523\times10^{-3}$ & $3.582\times 10^4$ & 265.5 & 1204  & $-3205$ \\[1ex]
   5    & 1   & 714.0 & $1.584\times10^{-3}$ & $1.439\times 10^5$ & 570.2 & 2864  & $-7438$ \\
   5    & 3   & 719.9 & $1.563\times10^{-3}$ & $1.464\times 10^5$ & 574.8 & 2893  & $-7511$ \\
   5    & num & 720.0 & $1.563\times10^{-3}$ & $1.464\times 10^5$ & 574.9 & 2894  & $-7513$ \\
   \bottomrule
  \end{tabular}
  \end{minipage}
 \end{table*}

 \subsection{Analytic Solution for Polytropes with $\bmath{\lowercase{n}=1}$}
  \subsubsection{The Zeroth Order: $\sigma^0$}
  We rewrite \eqref{Lane-Emden} for $n=1$, remembering that now $\tilde\mu=\mu$,
  \begin{equation}
   \left(\frac{\d^2}{\d \bar{r}^2}+ \frac{1}{\bar{r}}\frac{\d}{\d \bar{r}}\right)\mu_{00} + 2\pi\mu_{00}= 0
  \end{equation}
  and can immediately write down the general solution
  \begin{equation}
   \mu_{00} = C_1 J_0(\sqrt{2\pi} \bar{r}) + C_2 Y_0(\sqrt{2\pi} \bar{r}),
  \end{equation}
  where $J_n$ is a Bessel function (of the first kind) and $Y_n$ a Neumann function (also called a
  Bessel function of the second kind), see e.g.\ \citet*{PBM90b}. The condition
  $\mu(r=0,\chi)=\mu_\text c$ tells us that $C_1=1$
  and $C_2=0$. The first positive zero of $J_0$ determines value
  for $a$ from \eqref{rs}. We refer to the $k$th positive zero of the $n$th Bessel function as $j_{nk}$
  and can then write
  \begin{equation}
   \bar{a}:= a \bar{r}/r= j_{01}/\sqrt{2\pi} = 0.959\ldots
  \end{equation}
  The value for $a=\bar a\sqrt{K/G}$ is independent of the choice of $\mu_\text c$, which is not the case
  for other polytropic indices. This is due to an interesting invariance for $n=1$:
  if $U(\bmath x)$, $\mu(\bmath x)=K\sqrt{p(\bmath x)}$ and $\bmath v(\bmath x)$ are solutions to the Poisson
  and Euler equations, then so are $\alpha U(\bmath x)$, $\alpha \mu(\bmath x)=K \sqrt{\alpha^2 p(\bmath x)}$
  and $\sqrt{\alpha}\bmath v(\bmath x)$, where $\alpha$ is an arbitrary scaling factor.

 \subsubsection{The First Order: $\sigma^1$}
  The unknown quantities we have to solve for are $\mu_{10}(\bar{r})$, $\mu_{11}(\bar{r})$, $\beta_{10}$, $\beta_{11}$ and $\Omega_2$.
  From \eqref{Laplace_Euler}, one finds the differential equations
  \begin{align}
   &\left(\frac{\d^2}{\d \bar{r}^2}+ \frac{1}{\bar{r}}\frac{\d}{\d \bar{r}}\right)\mu_{10} + 2\pi\mu_{10}= 0 \intertext{and}
   &\left(\frac{\d^2}{\d \bar{r}^2}+ \frac{1}{\bar{r}}\frac{\d}{\d \bar{r}}\right)\mu_{11} + \left(2\pi-\frac{1}{\bar{r}^2}\right)\mu_{11}=
      \frac{1}{ \bar a}\frac{\d\mu_{00}}{\d \bar{r}}.
  \end{align}
  Considering only solutions that vanish at the point $\bar{r}=0$, so as to maintain our choice $\mu(0)=\mu_\text c$,
  we find
  \begin{align}
   \mu_{10} &= 0 \intertext{and}
   \mu_{11} &= C_3 J_1 + \frac{1}{2 j_{01}}\left(\sqrt{2\pi}\bar{r} J_0 - J_1\right),
  \end{align}
  where the argument of the Bessel functions is always $\sqrt{2\pi}\bar{r}$ unless otherwise specified.
  The requirement that the density vanish at the surface of the rings determines
  \begin{align}
   \beta_{10}=0
  \end{align}
  and relates the constant $C_3$ to the surface function
  \begin{equation}
   C_3 = \frac{1+2 j_{01}^2 \beta_{11}}{2j_{01}}.
  \end{equation}
  The constant $\beta_{11}$ is determined by stipulating that the centre of mass coincide with the
  point $(\varrho=b,z=0)$ as in \eqref{com}
  \begin{equation}
   \beta_{11} = \frac{4-j_{01}^2}{4j_{01}^2}.
  \end{equation}
  Recalling the definition $\lambda := \ln\frac{8}{\sigma}-2$, one finally obtains
  \begin{equation}
   \Omega_2 = \frac{2J_1(j_{01})(\lambda+1)}{j_{01}}
  \end{equation}
  from \eqref{int_Euler_surf}.

 \subsubsection{The Second Order: $\sigma^2$}
  To second order, the unknown quantities that have to be solved for are $\mu_{20}(\bar{r})$, $\mu_{21}(\bar{r})$, $\mu_{22}(\bar{r})$,
  $\beta_{20}$, $\beta_{21}$, $\beta_{22}$ and $\Omega_3$.

  The ODEs describing the mass density now read
  \begin{align}
  \begin{split}
   &\left(\frac{\d^2}{\d \bar{r}^2}+ \frac{1}{\bar{r}}\frac{\d}{\d \bar{r}}\right)\mu_{20} + 2\pi\mu_{20}= \pi\Omega_2
      \\& \qquad+ \frac{1}{2 \bar a}\left(\frac{\d\mu_{11}}{\d \bar{r}} - \frac{\mu_{11}}{\bar{r}}\right)
               + \frac{\bar{r}}{2 \bar a^2}\frac{\d\mu_{00}}{\d \bar{r}},
  \end{split}\\
   &\left(\frac{\d^2}{\d \bar{r}^2}+ \frac{1}{\bar{r}}\frac{\d}{\d \bar{r}}\right)\mu_{21} + \left(2\pi-\frac{1}{\bar{r}^2}\right)\mu_{21}-
      \frac{1}{ \bar a}\frac{\d\mu_{10}}{\d \bar{r}}=0\intertext{and}
   \begin{split}
   &\left(\frac{\d^2}{\d \bar{r}^2}+ \frac{1}{\bar{r}}\frac{\d}{\d \bar{r}}\right)\mu_{22} + \left(2\pi-\frac{4}{\bar{r}^2}\right)\mu_{22}=
      \\& \qquad \frac{1}{2 \bar a}\left(\frac{\d\mu_{11}}{\d \bar{r}} - \frac{\mu_{11}}{\bar{r}}\right) 
               + \frac{\bar{r}}{2 \bar a^2}\frac{\d\mu_{00}}{\d \bar{r}}.
   \end{split}
  \end{align}
  The solutions vanishing at $\bar{r}=0$ are
  \begin{align}
   \mu_{20} &= \frac{\Omega_2}{2}\left(1-J_0\right) +
      \frac{1}{8}\left[\frac{3\pi \bar{r}^2}{j_{01}^2}J_0 + \sqrt{2\pi}\bar{r} \left(\frac{1}{j_{01}^2}-\frac{1}{2}\right)J_1 \right], \\
   \mu_{21} &= C_4 J_1 \intertext{and}
  \begin{split}
   \mu_{22} &= C_5 J_2 
    + \frac{1}{4}\left[\left(\frac{5}{j_{01}^2} + \frac{3\pi \bar{r}^2}{2 j_{01}^2} -\frac{1}{2}\right)J_2 \right.\\
    & \qquad +\left. \frac{\sqrt{\pi}\bar{r}}{\sqrt{2}}\left(\frac{5}{j_{01}^2}-\frac{1}{2}\right)J_1\right].
   \end{split}
  \end{align}

 The constants $C_4$ and $C_5$ can be related to the surface function by requiring that $\mu(\rs)=0$ hold independently
 for the coefficients in front of $\cos\chi$ and $\cos 2\chi$. The result is
 \begin{align}
  C_4 &= j_{01} \beta_{21} \intertext{and}
  C_5 &= \frac{1}{2}\left(-j_{01}^2 \beta_{22} -\frac{j_{01}^2}{64}+\frac{3}{8} -\frac{11}{4j_{01}^2}\right).
 \end{align}
 Requiring the same of the coefficient in front of $\cos 0\chi$ gives
 \begin{equation}
  \beta_{20} = -\frac{4}{j_{01}^4} + \frac{\lambda+1}{j_{01}^2} -\frac{1}{64}.
 \end{equation}
 Evaluating \eqref{com} tells us that
 \begin{equation}
  \beta_{21} = 0 \Longrightarrow \mu_{21}=0.
 \end{equation}
 The values for the remaining constants follow from \eqref{int_Euler_surf}:
 \begin{align}
  \Omega_3= 0
 \end{align}
 and
 \begin{align}
  \beta_{22}  = \frac{1}{4j_{01}^4} + \frac{5(\lambda+3)}{2j_{01}^2} + \frac{1}{64}.
 \end{align}

\subsubsection{The Third Order: $\sigma^3$}

 The third order is the final one to be presented here, but the iterative scheme can
 be applied up to arbitrary order assuming that one is able to solve the differential
 equations for the mass density and perform the necessary integrals. The ODEs that result for this order are
 \begin{align}
   &\left(\frac{\d^2}{\d \bar{r}^2}+ \frac{1}{\bar{r}}\frac{\d}{\d \bar{r}}\right)\mu_{30} + 2\pi\mu_{30}= 0,\\
   \begin{split}
   &\left(\frac{\d^2}{\d \bar{r}^2}+ \frac{1}{\bar{r}}\frac{\d}{\d \bar{r}}\right)\mu_{31} + \left(2\pi-\frac{1}{\bar{r}^2}\right)\mu_{31}=
        \frac{3\bar{r}^2}{4 \bar a^3}\frac{\d\mu_{00}}{\d \bar{r}}
      \\& \quad  + \frac{3}{4 \bar a^2}\left(\bar{r}\frac{\d}{\d \bar{r}} + 1 \right)\mu_{11}
      + \frac{1}{ \bar a}\left[\frac{\d\mu_{20}}{\d \bar{r}} + \left(\frac{1}{2}\frac{\d}{\d \bar{r}}+ \frac{1}{\bar{r}}\right)\mu_{22}\right],
   \end{split}\\
   &\left(\frac{\d^2}{\d \bar{r}^2}+ \frac{1}{\bar{r}}\frac{\d}{\d \bar{r}}\right)\mu_{32} + 2\left(\pi -\frac{2}{\bar{r}^2}\right)\mu_{32}= 0
   \intertext{and}
   \begin{split}
   &\left(\frac{\d^2}{\d \bar{r}^2}+ \frac{1}{\bar{r}}\frac{\d}{\d \bar{r}}\right)\mu_{33} + \left(2\pi-\frac{9}{\bar{r}^2}\right)\mu_{33}=
       \frac{\bar{r}^2}{4 \bar a^3}\frac{\d\mu_{00}}{\d \bar{r}}
      \\& \quad + \frac{1}{4 \bar a^2}\left(\bar{r}\frac{\d}{\d \bar{r}} - 1 \right)\mu_{11}
       + \frac{1}{ \bar a}\left(\frac{1}{2}\frac{\d}{\d \bar{r}}- \frac{1}{\bar{r}}\right)\mu_{22}.
   \end{split}
 \end{align}
 The solutions to these equations obeying the requirement $\mu_{qk}(0)=0$ are
 \begin{align}
  \mu_{30}&=0,\\
  \begin{split}
   \mu_{31}&=
    \left(C_6
       -\frac {\left( \lambda+1 \right)J_1\!\left(j_{01} \right) }{2j_{01}^{2}}
       -{\frac {9}{64}} {\frac {\pi  \left( j_{01}^{2}-8\right) {\bar{r}}^{2}}{j_{01}^{3}}}\right.\\
     &-\left.\frac {j_{01}^{4} + 16j_{01}^{2}(5 \lambda +4) -16}{256 j_{01}^{3}} \right) J_1\\
     & + \left( \frac {\sqrt {\pi }\left( \lambda+1 \right) J_1\!\left(j_{01} \right)\bar{r} }{\sqrt{2}j_{01}^{2}}
        -{\frac {15}{32}} {\frac {\sqrt {2}{\pi }^{3/2}{\bar{r}}^{3}}{j_{01}^{3}}} \right.\\
     &+\left.\frac {\sqrt {2\pi }\left(j_{01}^{4} + 16j_{01}^{2}(5\lambda+4)-16
         \right) \bar{r}}{256 j_{01}^{3}} \right) J_2,
  \end{split}\\
  \mu_{32}&=C_7 J_2\intertext{and}
  \begin{split}
    \mu_{33}&=
    \frac{\pi  {\bar{r}}^{2}\left(16j_{01}^{2}(5\lambda+2)-40 \pi {\bar{r}}^{2}+272+j_{01}^{4}\right)}{512j_{01}^{3}} J_1 \\
    &+\left(-\frac {5{\pi }^{2}{\bar{r}}^{4}}{64 j_{01}^{3}}
      +\frac {\pi \left(j_{01}^{4}+8j_{01}^{2}(10\lambda+7)+16\right)\bar{r}^2}{512 j_{01}^{3}}\right.\\
    &-\left.\frac {3\left(j_{01}^{4}+16j_{01}^{2}(5\lambda + 3)+208\right)}
       {256 j_{01}^{3}} + C_8\right)J_3.
  \end{split}
 \end{align}

 The constants $\beta_{30}$, $C_6$, $C_7$ and $C_8$ are determined by requiring that
 $\mu(r=r_\text s)=0$ hold independently for the coefficients in front
 of the $\cos 0\chi$, $\cos \chi$, $\cos 2\chi$ and $\cos 3\chi$ terms. The result is
 \begin{align}
  \begin{split}
   \beta_{30}&= 0,\\
   C_6&=
   \left( \frac{\lambda+1}{4} -\frac{3\left( \lambda+1 \right) }{2j_{01}^{2}} \right) J_1\!\left(j_{01} \right)
     -{\frac {j_{01}^{3}}{512}}\\
   &\qquad + j_{01}   \left( \frac{5}{32}\lambda+\beta_{31}+\frac{39}{256} \right)\\
   &\qquad -\frac{ 14 \lambda + 9 }{32 j_{01}}
    +\frac { 40 \lambda + 37}{16 j_{01}^{3}},
  \end{split}\\
  C_7&=0\intertext{and}
  \begin{split}
   C_8&= \frac{1}{j_{01}^2-8}\Bigg[-{\frac {j_{01}^{5}}{1536}}
     - \left({\frac {5}{32}} \lambda +\beta_{33}+{\frac {77}{768}}\right) j_{01}^{3}\\
   & \quad + \left( {\frac {35}{16}} \lambda+{\frac {11}{8}} \right) j_{01}
     - \frac{1}{ {j_{01}}}\left( 10 \lambda+{\frac {181}{48}} \right)
   -{\frac {119}{6 j_{01}^{3}}}\Bigg].
  \end{split}
 \end{align}
 Equation \eqref{com} yields
 \begin{align}
   \beta_{31}= {\frac {9}{32}} \lambda + {\frac {31}{128}}
     - \frac{1}{ j_{01}^{2}}\left({\frac {13}{8}} \lambda + {\frac {5}{16}} \right)
     -\frac{1}{j_{01}^{4}}\left(\frac{5}{2} \lambda + 9\right)
 \end{align}
 and \eqref{int_Euler_surf} gives
 \begin{align}
  \beta_{32}&=0,\\
  \begin{split}
   \Omega_4&=-2 {\frac { \left(\lambda +1\right) ^{2} }{j_{01}^{2}}} J_1\!\left(j_{01} \right)^2
    + \Bigg[ \frac{1}{32}\left(\lambda +1 \right) j_{01}\\
   & \quad +{\frac{1}{j_{01}}}\left({\lambda}^{2}+\frac{3}{8} \lambda-\frac{1}{2}\right)\\
   & \quad + \frac{1}{j_{01}^3}\left( \frac{11}{2}\lambda + {\frac {17}{4}}\right)  \Bigg] J_1\!\left(j_{01}\right)
  \end{split}\intertext{and}
  \beta_{33}&={\frac {5}{64}}- \frac{1}{8 j_{01}^{2}}\left(5\lambda +9 \right)
    +\frac{1}{j_{01}^{4}} \left(\frac{5}{2}\lambda+2\right).
 \end{align}

 \subsubsection{Physical Parameters to the Third Order}

\begin{figure}
  \centerline{\includegraphics{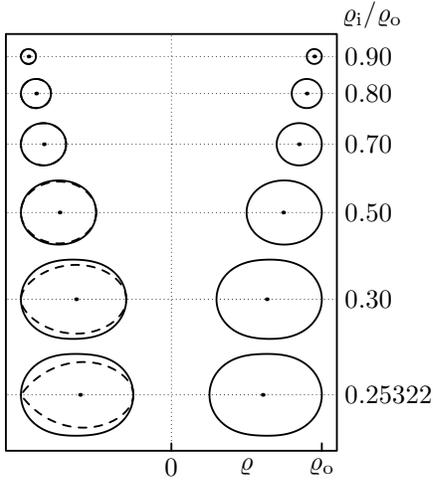}}
  \caption{Meridional cross-sections of polytropic rings ($n=1$) with varying radius ratio are shown to third order
           (solid lines) in comparison to numerical results (dashed lines) for the same radius
           ratio. At the value $\varrho_\text i/\varrho_\text o=0.25322\ldots$, the rings reach a mass-shedding
           limit, as is evident from the numerical cross-section.\label{fig_rs}}
 \end{figure}

 The shape of the rings to third order that results from equation \eqref{rs}
 is compared to numerical results of the corresponding radius ratio in Fig.~\ref{fig_rs}.
 For thin rings, the numerical and third order curves are indistinguishable. As the radius
 ratio is decreased, the numerical results show that the outer edge becomes pointier, right up
 to the mass-shedding limit for the value $\varrho_\text i/\varrho_\text o=0.25322\ldots$.
 For such a ring, a fluid particle rotating at the outer rim in the equatorial plane has a
 rotational frequency equal to the Kepler frequency, meaning that it is kept in balance by
 the gravitational and centrifugal forces alone -- the force arising from the pressure gradient
 vanishes. The shape of the ring with the cusp that forms for mass-shedding configurations
 is not well represented by a small number of terms in our Fourier series.

 Using the results of the last subsection, we write down expressions for various
 physical parameters and can use them to verify that the virial
 identity \eqref{VI} is satisfied to each order in $\sigma$.
 Up to and including third order one finds
 \begin{align}
   \begin{split}
   &\bar M= \,
     {\frac {\sqrt {2\pi }j_{01}^{2}J_1\!\left( j_{01} \right) }{\sigma}}\\
   &\qquad +\bigg[\frac{\sqrt {2\pi }}{64} \left(j_{01}^{4} +  28 j_{01}^{2}+32 j_{01}^{2}\lambda-16 \right)
            J_1\!\left( j_{01} \right)\\
   &\qquad  -\sqrt {2\pi } j_{01}  \left( \lambda+1 \right)
             J_1\!\left(j_{01} \right)^2  \bigg]  \sigma,
  \end{split}\\
  \begin{split}
   &\bar J= \,
       \sqrt {{\frac {J_1\!\left(j_{01} \right)\left( \lambda+1 \right) }{ j_{01}}}}J_1\!\left(j_{01} \right)j_{01}^{2}
     \Bigg\{\frac{j_{01}^{2}}{ {\sigma}^2}\\
     &\qquad     -\frac{3}{2}\left( \lambda+1 \right)j_{01} J_1\!\left(j_{01} \right)\\
     &\qquad + {\frac {1}{128\left( \lambda+1 \right)}}\big[3 j_{01}^{4}(\lambda+1)\\
     &\qquad +j_{01}^{2}(96 {\lambda}^{2}+324 \lambda+232) -624\lambda-664 \big]\Bigg\},
  \end{split}\\
  \begin{split}
   &\bar P=\,
      \frac{\sqrt {2\pi}}{2}j_{01} J_1\!\left(j_{01} \right)^2\Bigg\{
      \frac{j_{01}^{2}}{\sigma}\\
      &\qquad+\frac {1}{320}
       \Big[ -640 j_{01} \left( \lambda+1 \right)J_1\!\left(j_{01} \right)\\
      &\qquad+10\left( j_{01}^{4}-8 j_{01}^{2}+128\lambda+136 \right) \Big] \sigma \Bigg\},
  \end{split}\\
  \begin{split}
   &\bar T=\,
      \frac{\sqrt {2\pi}}{2} J_1\!\left(j_{01} \right)  ^2 { j_{01}}\Bigg\{
      \frac {1}{\sigma} \left( \lambda+1 \right) j_{01}^{2}\\
    &\qquad +\Big[ - 2\left( \lambda+1 \right) ^{2}J_1\!\left( j_{01} \right)
       { j_{01}}+{\frac {1}{32}}\Big(j_{01}^{4}(\lambda+1)\\
    &\qquad + 2j_{01}^{2}(8\lambda+9)(2\lambda+3)-112\lambda-132\Big)\Big] \sigma \Bigg\}
  \end{split}\intertext{and}
  \begin{split}
   &\bar W=\,
     \sqrt{2\pi}j_{01}J_1\!\left(j_{01} \right)^2\Bigg\{
     -{\frac{j_{01}^{2}}{2\sigma} \left( 2 \lambda+5 \right)}\\
    &\qquad +\Bigg[ { j_{01}} \left( 2 \lambda+5 \right)  \left( \lambda+1 \right)J_1\!\left(j_{01} \right)\\
    &\qquad -{\frac {1}{64}}\Big(j_{01}^{4}(2 \lambda+5)+ 4 j_{01}^{2}(16{\lambda}^{2}+42 \lambda+21)\\
    &\qquad\qquad+160\lambda +144\Big)\Bigg] \sigma\Bigg\}.
  \end{split}
 \end{align}
 In the derivation of the above expressions for $\bar P$ and $\bar W$, we have made use
 of the identity
 \begin{align}\label{F_identity}
   20\, _2\!F_3\!\left(\frac{3}{2},\frac{3}{2};2,2,\frac{5}{2};-j_{01}^{2} \right)
   = 3j_{01}^{2}\, _2\!F_3\!\left(\frac{5}{2},\frac{5}{2};3,3,\frac{7}{2};-j_{01}^{2} \right)
 \end{align}
 for the Gauss hypergeometric function
 \begin{align}
  \begin{split}
 & _p\!F_q(a_1,a_2,\ldots,a_p;b_1,b_2,\ldots,b_q;z)\\
 &\qquad :=\sum_{k=0}^\infty
    \frac{{(a_1)}_k \cdot{(a_2)}_k \cdots {(a_p)}_k}{{(b_1)}_k \cdot{(b_2)}_k \cdots {(b_q)}_k}\frac{z^k}{k!}
  \end{split}\intertext{with the Pochhammer bracket}
 & {(a)}_k:= a(a+1)\cdots(a+k-1), \qquad {(a)}_0:=1.\nonumber
 \end{align}
 A proof of \eqref{F_identity} can be found in Appendix~\ref{app:F}.
 \par

\begin{figure}
  \centerline{\includegraphics{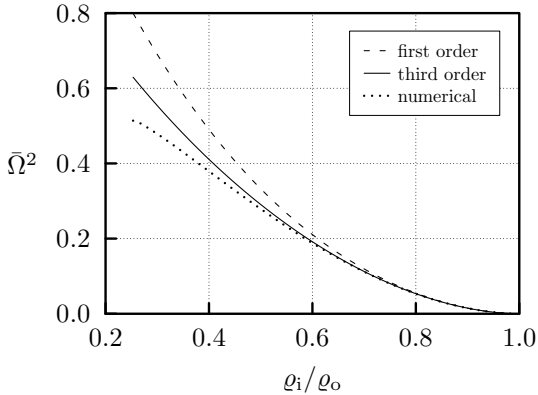}}
  \caption{The square of the dimensionless angular velocity is plotted versus
           the radius ratio for rings with polytropic index $n=1$.\label{Om2_n_poly_1}}
 \end{figure}

 In order to gauge the accuracy of the expressions listed above, some of them are
 plotted to first and third order in comparison to numerical values in Figs~\ref{Om2_n_poly_1}--\ref{J_n_poly_1}.
 The accuracy of the numerical values is high enough so as to render the corresponding curve
 indistinguishable from the `correct' one and is plotted in its entirety, i.e.\ from the thin
 ring limit right up to the mass-shedding limit.
 The curves to first and third order were drawn by taking the expression for $\bar{\varrho}_\text i$,
 $\bar{\varrho}_\text o$, $\bar{\Omega}^2$, $\bar{M}$ and $\bar{J}$ to first and third order respectively,
 inserting a numerical value for $\sigma$ and then taking the appropriate combination of these numbers.
 One finds in all three plots that the third order brings a marked improvement as compared to the first
 one, but that the behaviour near the mass-shedding limit is not particularly well represented.
 \begin{figure}
  \centerline{\includegraphics{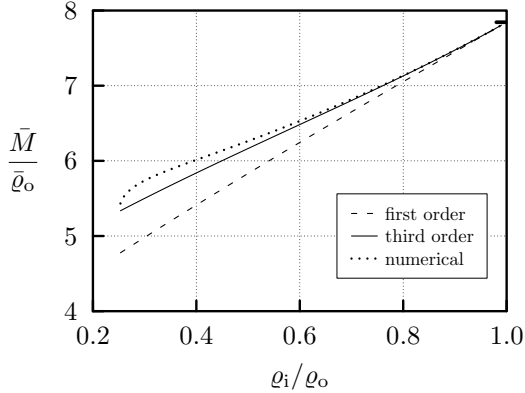}}
  \caption{The dimensionless mass divided by the outer radius is plotted versus
           the radius ratio for rings with polytropic index $n=1$. This quantity tends to
           the value $\bar{M}/\bar{\varrho}_\text o=2\pi j_{01} J_1(j_{01})=7.84\ldots$
           in the thin ring limit, which is marked by a tick.}
 \end{figure}
 \begin{figure}
  \centerline{\includegraphics{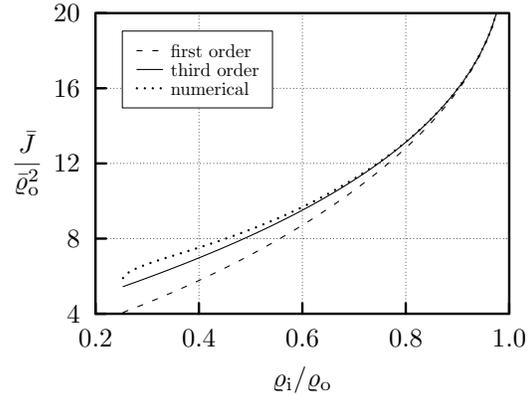}}
  \caption{The dimensionless angular momentum divided by the square of the outer radius is plotted versus
           the radius ratio for rings with polytropic index $n=1$. This quantity as a function of
           radius ratio tends logarithmically to infinity.\label{J_n_poly_1}}
 \end{figure}

 \section{A Limit of Infinite Polytropic Index}\label{inf_pol}

\begin{figure}
  \centerline{\includegraphics{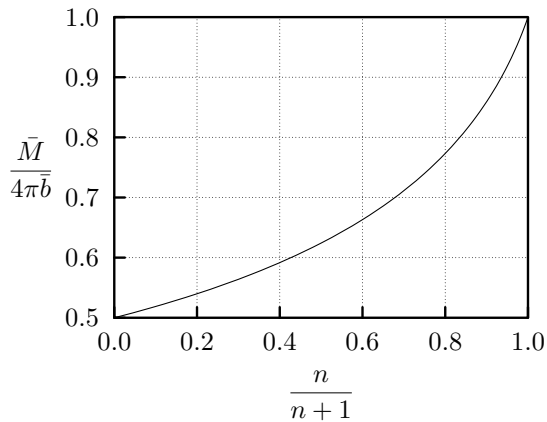}}
  \caption{The dimensionless mass divided by $4\pi\bar b$ in the thin ring limit is plotted versus $n/(n+1)$ over
           the whole range of polytropic indices $n\in [0,\infty)$. The points for $n=0$,
           $n=1$ and $n\to\infty$ are known analytically and the remainder of the curve was
           generated by solving the equation for $\tilde\mu_{00}$ numerically and making use
           of equation \eqref{mass_h00}.\label{M_4pic}}
 \end{figure}

 As $n$ tends to infinity, the polytropic equation \eqref{poly} shows us that pressure and
 density are proportional
 \begin{align}\label{isothermal} p=K\mu, \end{align}
 a case sometimes referred to as `isothermal' because such an equation holds for
 an ideal gas at constant temperature.
 Inserting this into equation \eqref{Laplace_Euler} at leading
 order and again using the dimensionless coordinate $\bar{r}$ yields
 \begin{align}
  4\pi \bar{r}\mu_{00} + \frac{\d}{\d \bar{r}}\left(\bar{r} \frac{\d}{\d \bar{r}}\ln\mu_{00} \right)=0.
 \end{align}
 The solution to this equation with our normalization $\mu_{00}(\bar{r}=0)=1$ reads
 \begin{align}\label{mu_inf}
  \mu_{00} = \left(\frac\pi2\bar{r}^2+1\right)^{-2}.
 \end{align}
 The density and pressure fall to zero as $\bar{r}\to\infty \Leftrightarrow r\to\infty$.
 Integrating over the density to calculate the normalized mass, one finds to leading order
 \begin{align}
  \bar M = 4\pi^2 \bar b \int_0^\infty \mu_{00} \bar{r}\, \d \bar{r} = 4\pi\bar b,
 \end{align}
 which can also be read off from equation \eqref{eq:P_M2} directly, by making use
 of $\bar M=\bar P$, which is self-evident upon taking \eqref{isothermal} into account.
 In Fig.~\ref{M_4pic}, the behaviour of $\bar M/4\pi\bar b$ can be followed from
 the homogeneous case, $n=0$, right up to the isothermal limit $n\to\infty$.

 Making use of \eqref{mu_inf}, we find that
 \begin{align}
   \lim_{n\to\infty} g=0,
 \end{align}
 where $g$ was defined in \eqref{def_g}. It thus follows from \eqref{g_minus_beta11} that
 \begin{align}
   \lim_{n\to\infty}\left( \beta_{11} +\frac{n}{8}\right)= \frac{1}{8}
 \end{align}
 as already suggested by the results of Table~\ref{beta11_Om2}. We can then see
 that the specific kinetic energy $\bar a^2\Omega_2$ tends to infinity such that
 for fixed $\lambda$
 \begin{align}
 \lim_{n\to\infty} \frac{2\pi\bar a^2\Omega_2}{n+4\lambda} = 1.
 \end{align}

 From the fact that $|\beta_{11}|$ tends to infinity, we can conclude that the range of $\sigma$ values
 for which the first order provides a good approximation shrinks to the point $\sigma=0$.
 In general, for a given equation of state, there exist ring solutions over an interval of
 radius ratios ranging from 1 in the thin ring limit down to a minimal value at the
 mass-shedding limit. The numerical values presented in Table~\ref{mass-shed_vs_n}
 demonstrate that this minimal value grows for increasing $n$.
 \begin{table}
  \centering
  \caption{The radius ratio of the mass-shedding ring for various polytropic indices $n$. The value
           $\varrho_\text i/\varrho_\text o=0$ indicates the transition from toroidal to spheroidal
           topologies, which only exists for $n\la 0.36$.\label{mass-shed_vs_n}}
  \tabcolsep1ex
  \begin{tabular}{cccccccccc}\toprule
    $n$ & 0.36 & 1 & 2 & 3 & 4 & 5 & 6 & 7 & 8  \\
    $\varrho_\text i/\varrho_\text o$ & 0 &  0.25 &  0.37 &  0.44 &  0.49 &  0.53 &  0.56  &  0.58 &  0.60\\
   \bottomrule
  \end{tabular}
 \end{table}
 The value for $\varrho_\text i/\varrho_\text o$
 at which one reaches the mass-shedding limit presumably tends to 1 as $n$ tends to infinity.
 This would imply that the isothermal thin ring limit presented above and even to first order by
 \citet{Ostriker64b}, i.e.\ in which the cross-section of the ring tends to that of a circle, is not unique. Analytic
 work including a family of isothermal thin ring limits and containing the mass-shedding limit
 will be presented elsewhere.

 \section{Concluding Remarks}

 In their work on homogeneous rings, Poincar\'e and Kowalewsky, whose
 results disagreed to first order, both had made mistakes as Dyson has shown. His
 result to fourth order is also erroneous as we point out in the appendix. It thus seems
 particularly worthwhile to test the correctness of the solutions presented here.
 For one thing, we ensured that the transition condition
 \begin{align}
 \nabla U_\text{in}|_\text{s}=\nabla U_\text{out}|_\text{s}
 \end{align}
 is fulfilled up to the appropriate order in $\sigma$.
 Furthermore we tested that the virial theorem \eqref{VI}
 is fulfilled for each order in $\sigma$.
 \par
 Please note that one has to be careful in interpreting the results for the thin ring limit.
 For example, one might think that the squared angular velocity vanishes like $\sigma^2\ln\sigma$.
 This is true for the dimensionless quantity $\Omega^2/G\mu_\text c$, but need not be true for the
 squared angular velocity itself. If we fix the `size' $b$ and the mass $M$ of the ring
 in that limit, then the cross-section shrinks to a point ($a=\sigma b$).
 With \eqref{mass_h00} we can conclude that $\mu_\text c\propto\sigma^{-2}$
 and therefore
 $\Omega^2\propto\ln\sigma$, which means that $\Omega^2$ and hence
 the velocity of a fluid element tend to infinity.
 \par
 Relativistic rings, including the thin ring limit, were studied in \citet{AKM03c},
 \citet{Ansorgetal04} and \citet{FHA05}.
 From the perspective of General Relativity, the Newtonian theory constitutes
 a good approximation when certain conditions are fulfilled. For one thing,
 typical velocities must be small compared to the speed of light $c$ and
 for another $|U|\ll c^2$ must hold. We just saw, however, that for rings
 of finite extent and mass, the velocities grow unboundedly in the thin
 ring limit. The same holds for $U_\text s$ as well, see \eqref{Us_general}.
 This means that the Newtonian theory of gravity is not
 appropriate to describe this subtle limit itself, since one cannot expect
 it to be a good approximation to General Relativity.
 It is remarkable that the approximation about the point $\sigma=0$ is
 nevertheless so successful.

 \section*{Acknowledgments}
   It is a pleasure to thank Professor R.\ Meinel for fruitful discussions.
   The authors are also grateful to Professor J.\ Ostriker for pointing
   out his work on this subject to us.
   Many of the computations in this paper made use of Maple\texttrademark. Maple is a
   trademark of Waterloo Maple Inc.
   This research was funded in part by the Deutsche Forschungsgemeinschaft
   (SFB/TR7--B1).

 \bibliographystyle{mn2e}
 \bibliography{Reflink}

 \appendix
 \section{An Identity Relating Hypergeometric to Bessel Functions}\label{app:F}
 In order to prove the identity \eqref{F_identity}, we prove the more general
 identity
 \begin{align}\label{app:id}
  \begin{split}
   & \frac{3}{40}z\, _2\!F_3\left(\frac{5}{2},\frac{5}{2};3,3,\frac{7}{2};-z \right)
    - \frac{1}{2}\, _2\!F_3\left(\frac{3}{2},\frac{3}{2};2,2,\frac{5}{2};-z \right)\\
   &\qquad = \frac{\d}{\d z}\left[J_0(\sqrt z)^2 \right],
  \end{split}
 \end{align}
 for an arbitrary complex number $z$, from which \eqref{F_identity} follows immediately.

 We begin by using the differentiation properties of the hypergeometric functions, e.g.\
 7.2.3.47 in \citet{PBM90b}, to write
 \begin{align}
  \begin{split}
   & \frac{3}{40}\,\, _2\!F_3\left(\frac{5}{2},\frac{5}{2};3,3,\frac{7}{2};-z \right) z
    - \frac{1}{2}\,\, _2\!F_3\left(\frac{3}{2},\frac{3}{2};2,2,\frac{5}{2};-z \right)\\
   &\qquad = \frac{\d}{\d z}\Bigg[\,_2\!F_3\left(\frac{1}{2},\frac{1}{2};1,1,\frac{3}{2};-z \right)\\ &\qquad\qquad\qquad
                  - \frac{z}{3}\,\,_2\!F_3\left(\frac{3}{2},\frac{3}{2};2,2,\frac{5}{2};-z \right) \Bigg].
  \end{split}
 \end{align}
 With the integral identity 7.2.3.11, the term to be differentiated can
 be written as
 \begin{align}
  \begin{split}
   & _2\!F_3\left(\frac{1}{2},\frac{1}{2};1,1,\frac{3}{2};-z \right)
                  - \frac{z}{3}\,\,_2\!F_3\left(\frac{3}{2},\frac{3}{2};2,2,\frac{5}{2};-z \right)\\
   & \quad = \frac{1}{\pi}\int_0^1\!\!\!\int_0^1 \frac{J_0(2\sqrt{t_1 t_2 z})}{2\sqrt{t_1 t_2 (1-t_1)}}
            -\frac{z J_1(2\sqrt{t_1 t_2 z})}{2\sqrt{z(1-t_1)}} \,\d t_1 \, \d t_2,
  \end{split}
 \end{align}
 where we have made use of the identity (e.g.\ 7.13.1.1 in \citealt{PBM90b})
 \begin{align}
  _0\!F_1(b,-z) = \Gamma(b) z^{(1-b)/2} J_{b-1}(2\sqrt z).
 \end{align}
 The above double integral yields
 \begin{align}
  \begin{split}
   & \frac{1}{\pi}\int_0^1\!\!\!\int_0^1 \frac{J_0(2\sqrt{t_1 t_2 z})}{2\sqrt{t_1 t_2 (1-t_1)}}
            -\frac{z J_1(2\sqrt{t_1 t_2 z})}{2\sqrt{z(1-t_1)}} \,\d t_1 \, \d t_2\\
   & \qquad =\frac{1}{\pi}\int_0^1 \frac{J_0(2\sqrt{t_1 z})}{t_1 (1-t_1)} \d t_1 = J_0(\sqrt z)^2,
 \end{split}
 \end{align}
 thereby proving \eqref{app:id}.

\section{Further coefficients}

In Tables~\ref{tab_omega}--\ref{tab_alpha}, coefficients for $\Omega_i$, $\beta_{ik}$, $U_{ik}(y)$ and $\alpha_{li}$ 
for homogeneous rings are given.

\begin{table}
 \renewcommand{\arraystretch}{1.8}\renewcommand{\baselinestretch}{1.4}
 \centering
\caption{Coefficients $\Omega_i$ for homogeneous rings up to the order $q=9$ ($i=2,3,\ldots,10$).\label{tab_omega}}
  \input{omega.tex}
 \end{table}

\begin{landscape}
\begin{table}
\renewcommand{\arraystretch}{1.8}\renewcommand{\baselinestretch}{1.4}
\caption{Coefficients $\beta_{ik}$ for homogeneous rings up to the order $q=9$. The bold-faced type indicates that these terms
are incorrect in \citet{Dyson92}.\label{tab_beta}}
 \input{beta.tex}
 \end{table}
\end{landscape}

 \begin{landscape}
 \begin{table}
 \renewcommand{\arraystretch}{1.8}\renewcommand{\baselinestretch}{1.4}
 \caption{Coefficients $U_{ik}(y)$ for homogeneous rings up to the order $q=7$.\label{tab_Uinnen}}
  \input{Uinnen.tex}
  \end{table}
 \end{landscape}

 \begin{landscape}
 \begin{table}
 \renewcommand{\arraystretch}{1.8}\renewcommand{\baselinestretch}{1.4}
 \caption{Coefficients $\alpha_{li}$ for homogeneous rings up to the order $q=8$.\label{tab_alpha}}
  \input{alpha.tex}
  \end{table}
 \end{landscape}

 \label{lastpage}

\end{document}

%% file: omega.tex
\begin{tabular}{c|*{1}{>{\footnotesize\PBS\raggedright}p{7cm}}}
$i$ & $\Omega_i$ \\\hline2 & $\lambda+\frac{3}{4}
$\\3 & $0
$\\4 & $-\frac{1}{8}\,\lambda-{\frac {19}{96}}
$\\5 & $0
$\\6 & ${\frac {25}{128}}\,{\lambda}^{3}+{\frac {365}{1536}}\,{\lambda}^{2}-{
\frac {2345}{18432}}\,\lambda-{\frac {8989}{73728}}
$\\7 & $0
$\\8 & ${\frac {25}{64}}\,{\lambda}^{4}+{\frac {40235}{98304}}\,{\lambda}^{3}-
{\frac {134255}{294912}}\,{\lambda}^{2}-{\frac {36493505}{56623104}}\,
\lambda-{\frac {34831813}{226492416}}
$\\9 & $0
$\\10 & ${\frac {134925}{131072}}\,{\lambda}^{5}+{\frac {2797535}{1572864}}\,{
\lambda}^{4}-{\frac {50072105}{169869312}}\,{\lambda}^{3}-{\frac {
1021727845}{509607936}}\,{\lambda}^{2}-{\frac {104581877693}{
97844723712}}\,\lambda-{\frac {49377918425}{391378894848}}
$\\\end{tabular}

%% file: beta.tex
\begin{tabular}{c|*{9}{>{\footnotesize\PBS\raggedright}p{2.00cm}}}
{\Large $ _i\!\!\diagdown\!\!^k$} & 1 & 2 & 3 & 4 & 5 & 6 & 7 & 8 & 9  \\  \hline 
1 & $0
$&---&---&---&---&---&---&---&---\\2 & $0
$&$\frac{5}{8}\,\lambda+{\frac {35}{96}}
$&---&---&---&---&---&---&---\\3 & $0
$&$0
$&${\frac {5}{128}}\,\lambda-{\frac {35}{3072}}
$&---&---&---&---&---&---\\4 & $0
$&$\frac{5}{8}\,{\lambda}^{2}+\mathbf{{\frac {95}{128}}\,\lambda+{\frac {1145}{9216}}}
$&$0
$&${\frac {75}{256}}\,{\lambda}^{2}+\mathbf{{\frac {815}{2304}}\,\lambda+{\frac {
5089}{55296}}}
$&---&---&---&---&---\\5 & $-{\frac {25}{1024}}\,{\lambda}^{2}-{\frac {175}{24576}}\,\lambda+{
\frac {1225}{294912}}
$&$0
$&${\frac {15}{256}}\,{\lambda}^{2}+{\frac {4955}{147456}}\,\lambda-{
\frac {62141}{3538944}}
$&$0
$&${\frac {25}{512}}\,{\lambda}^{2}+{\frac {75}{4096}}\,\lambda-{\frac {
14455}{1179648}}
$&---&---&---&---\\6 & $0
$&${\frac {5185}{4096}}\,{\lambda}^{3}+{\frac {110515}{49152}}\,{\lambda}
^{2}+{\frac {853225}{884736}}\,\lambda+{\frac {34487}{1327104}}
$&$0
$&${\frac {75}{128}}\,{\lambda}^{3}+{\frac {116545}{110592}}\,{\lambda}^{
2}+{\frac {308395}{589824}}\,\lambda+{\frac {17892169}{318504960}}
$&$0
$&${\frac {625}{4096}}\,{\lambda}^{3}+{\frac {169625}{589824}}\,{\lambda}
^{2}+{\frac {1097461}{7077888}}\,\lambda+{\frac {7327349}{339738624}}
$&---&---&---\\7 & $-{\frac {125}{2048}}\,{\lambda}^{3}-{\frac {76325}{1179648}}\,{\lambda
}^{2}+{\frac {78175}{28311552}}\,\lambda+{\frac {1313095}{169869312}}
$&$0
$&${\frac {1325}{8192}}\,{\lambda}^{3}+{\frac {109915}{442368}}\,{\lambda
}^{2}+{\frac {863633}{18874368}}\,\lambda-{\frac {105833029}{
4076863488}}
$&$0
$&${\frac {125}{1024}}\,{\lambda}^{3}+{\frac {9815}{73728}}\,{\lambda}^{2
}-{\frac {472843}{56623104}}\,\lambda-{\frac {32985941}{1509949440}}
$&$0
$&${\frac {375}{8192}}\,{\lambda}^{3}+{\frac {4825}{98304}}\,{\lambda}^{2
}-{\frac {34745}{9437184}}\,\lambda-{\frac {268795}{28311552}}
$&---&---\\8 & $0
$&${\frac {11935}{4096}}\,{\lambda}^{4}+{\frac {15693265}{2359296}}\,{
\lambda}^{3}+{\frac {387327415}{84934656}}\,{\lambda}^{2}+{\frac {
1967050099}{2717908992}}\,\lambda-{\frac {21240691225}{195689447424}}
$&$0
$&${\frac {48175}{32768}}\,{\lambda}^{4}+{\frac {299613275}{84934656}}\,{
\lambda}^{3}+{\frac {86354441}{31850496}}\,{\lambda}^{2}+{\frac {
163195481003}{244611809280}}\,\lambda+{\frac {61946667647}{
14676708556800}}
$&$0
$&${\frac {1875}{4096}}\,{\lambda}^{4}+{\frac {1994875}{1769472}}\,{
\lambda}^{3}+{\frac {1544984219}{1698693120}}\,{\lambda}^{2}+{\frac {
26196760631}{101921587200}}\,\lambda+{\frac {523810097561}{
34245653299200}}
$&$0
$&${\frac {21875}{262144}}\,{\lambda}^{4}+{\frac {2069375}{9437184}}\,{
\lambda}^{3}+{\frac {7862435}{42467328}}\,{\lambda}^{2}+{\frac {
6142108445}{114152177664}}\,\lambda+{\frac {968695327}{342456532992}}
$&---\\9 & $-{\frac {26875}{131072}}\,{\lambda}^{4}-{\frac {10803575}{28311552}}\,
{\lambda}^{3}-{\frac {73205885}{452984832}}\,{\lambda}^{2}+{\frac {
245384393}{10871635968}}\,\lambda+{\frac {5148689159}{391378894848}}
$&$0
$&${\frac {58565}{131072}}\,{\lambda}^{4}+{\frac {317810795}{339738624}}
\,{\lambda}^{3}+{\frac {4061560685}{8153726976}}\,{\lambda}^{2}-{
\frac {3261000013}{97844723712}}\,\lambda-{\frac {1033257753043}{
23482733690880}}
$&$0
$&${\frac {3125}{8192}}\,{\lambda}^{4}+{\frac {20750915}{28311552}}\,{
\lambda}^{3}+{\frac {738672683}{2264924160}}\,{\lambda}^{2}-{\frac {
180456831247}{3261490790400}}\,\lambda-{\frac {13297677452861}{
365286968524800}}
$&$0
$&${\frac {2625}{16384}}\,{\lambda}^{4}+{\frac {2597725}{9437184}}\,{
\lambda}^{3}+{\frac {112005563}{1358954496}}\,{\lambda}^{2}-{\frac {
117378627479}{2283043553280}}\,\lambda-{\frac {173703027451}{
9132174213120}}
$&$0
$&${\frac {625}{16384}}\,{\lambda}^{4}+{\frac {72125}{1048576}}\,{\lambda
}^{3}+{\frac {4955105}{226492416}}\,{\lambda}^{2}-{\frac {79046365}{
5435817984}}\,\lambda-{\frac {1636799939}{260919263232}}
$\\\end{tabular}

%% file: Uinnen.tex
\begin{tabular}{c|*{1}{>{\footnotesize\PBS\raggedright}p{3.31cm}}
                  *{1}{>{\footnotesize\PBS\raggedright}p{3.20cm}}
                  *{1}{>{\footnotesize\PBS\raggedright}p{2.78cm}}
                  *{1}{>{\footnotesize\PBS\raggedright}p{2.94cm}}
                  *{1}{>{\footnotesize\PBS\raggedright}p{2.31cm}}
                  *{1}{>{\footnotesize\PBS\raggedright}p{1.83cm}}
                  *{1}{>{\footnotesize\PBS\raggedright}p{1.57cm}}
                  *{1}{>{\footnotesize\PBS\raggedright}p{1.57cm}}
}
{\Large $ _i\!\!\diagdown\!\!^k$} & 0 & 1 & 2 & 3 & 4 & 5 & 6 & 7  \\  \hline
0 & $2\,\lambda +5 -{y}^{2}
$&---&---&---&---&---&---&---\\
1 & $0
$&$\big( \lambda+1 \big)y -\frac{1}{4}\,{y}^{3}
$&---&---&---&---&---&---\\
2 & $-\frac{1}{8}\,\lambda-{\frac {7}{32}}+ \big( \frac{1}{4}\,\lambda+\frac{1}{4} \big) {y}^{2}
-{\frac {3}{32}}\,{y}^{4}
$&$0
$&$ \big( \lambda+{\frac {31}{48}} \big){y}^{2} -{\frac {5}{48}}\,{y}^
{4}
$&---&---&---&---&---\\
3 & $0
$&$ \big( -{\frac {13}{32}}\,\lambda-{\frac {65}{192}} \big)y +
 \big( {\frac {7}{16}}\,\lambda+{\frac {67}{192}} \big) {y}^{3}-{
\frac {15}{128}}\,{y}^{5}
$&$0
$&$\big( {\frac {15}{64}}\,\lambda+{\frac {175}{1536}} \big){y}^{3} -
{\frac {35}{768}}\,{y}^{5}
$&---&---&---&---\\
4 & ${\frac {25}{64}}\,{\lambda}^{3}+{\frac {295}{384}}\,{\lambda}^{2}+{
\frac {3457}{9216}}\,\lambda+{\frac {19}{576}}+ \big( -{\frac {13}{
128}}\,\lambda-{\frac {65}{768}} \big) {y}^{2}+ \big( {\frac {21}{
128}}\,\lambda+{\frac {67}{512}} \big) {y}^{4}-{\frac {25}{512}}\,{y
}^{6}
$&$0
$&$ \big( \frac{5}{8}\,{\lambda}^{2}+{\frac {311}{512}}\,\lambda+{\frac {
269}{36864}} \big){y}^{2} + \big( {\frac {185}{768}}\,\lambda+{\frac {3205
}{18432}} \big) {y}^{4}-{\frac {35}{512}}\,{y}^{6}
$&$0
$&$\big( {\frac {455}{4608}}\,\lambda+{\frac {4459}{110592}}
 \big){y}^{4}  -{\frac {21}{1024}}\,{y}^{6}
$&---&---&---\\
5 & $0
$&$\big( {\frac {25}{128}}\,{\lambda}^{3}-{\frac {355}{1536}}\,{
\lambda}^{2}-{\frac {20005}{36864}}\,\lambda-{\frac {11081}{73728}}
 \big)y + \big( {\frac {5}{32}}\,{\lambda}^{2}+{\frac {155}{2048}}\,
\lambda-{\frac {9091}{147456}} \big) {y}^{3}+ \big( {\frac {465}{
2048}}\,\lambda+{\frac {2855}{16384}} \big) {y}^{5}-{\frac {875}{
12288}}\,{y}^{7}
$&$0
$&$\big( {\frac {15}{256}}\,{\lambda}^{2}+{\frac {25}{73728}}\,
\lambda-{\frac {100015}{1769472}} \big){y}^{3} + \big( {\frac {4795}{36864
}}\,\lambda+{\frac {76223}{884736}} \big) {y}^{5}-{\frac {315}{8192}
}\,{y}^{7}
$&$0
$&$\big( {\frac {805}{18432}}\,\lambda+{\frac {26579}{1769472}}
 \big){y}^{5} -{\frac {77}{8192}}\,{y}^{7}
$&---&---\\
6 & ${\frac {25}{32}}\,{\lambda}^{4}+{\frac {95435}{49152}}\,{\lambda}^{3}+
{\frac {862135}{589824}}\,{\lambda}^{2}+{\frac {7959115}{28311552}}\,
\lambda-{\frac {551045}{28311552}}+ \big( {\frac {25}{512}}\,{\lambda
}^{3}-{\frac {355}{6144}}\,{\lambda}^{2}-{\frac {20005}{147456}}\,
\lambda-{\frac {11081}{294912}} \big) {y}^{2}+ \big( {\frac {15}{
256}}\,{\lambda}^{2}+{\frac {465}{16384}}\,\lambda-{\frac {9091}{
393216}} \big) {y}^{4}+ \big( {\frac {775}{8192}}\,\lambda+{\frac {
14275}{196608}} \big) {y}^{6}-{\frac {6125}{196608}}\,{y}^{8}
$&$0
$&$\big( {\frac {335}{256}}\,{\lambda}^{3}+{\frac {9895}{4608}}
\,{\lambda}^{2}+{\frac {440125}{589824}}\,\lambda-{\frac {1755973}{
42467328}} \big){y}^{2} + \big( {\frac {245}{3072}}\,{\lambda}^{2}+{\frac 
{9325}{294912}}\,\lambda-{\frac {31315}{786432}} \big) {y}^{4}+
 \big( {\frac {455}{3072}}\,\lambda+{\frac {32011}{294912}} \big) {
y}^{6}-{\frac {1575}{32768}}\,{y}^{8}
$&$0
$&$\big( {\frac {125}{6912}}\,{\lambda}^{2}-{\frac {2765}{
1769472}}\,\lambda-{\frac {1824697}{79626240}} \big){y}^{4} + \big( {
\frac {10241}{147456}}\,\lambda+{\frac {1504909}{35389440}} \big) {y
}^{6}-{\frac {693}{32768}}\,{y}^{8}
$&$0
$&$\big( {\frac {17549}{884736}}\,\lambda+{\frac {243479}{
42467328}} \big){y}^{6} -{\frac {143}{32768}}\,{y}^{8}
$&---\\
7 & $0
$&$\big( {\frac {25}{64}}\,{\lambda}^{4}-{\frac {40645}{98304}}\,{
\lambda}^{3}-{\frac {2042065}{1179648}}\,{\lambda}^{2}-{\frac {
58120985}{56623104}}\,\lambda-{\frac {39130979}{339738624}} \big)y +
 \big( {\frac {745}{2048}}\,{\lambda}^{3}+{\frac {36385}{73728}}\,{
\lambda}^{2}+{\frac {200065}{2359296}}\,\lambda-{\frac {6542965}{
169869312}} \big) {y}^{3}+ \big( {\frac {645}{8192}}\,{\lambda}^{2}
+{\frac {27925}{786432}}\,\lambda-{\frac {645475}{18874368}} \big) {
y}^{5}+ \big( {\frac {85225}{589824}}\,\lambda+{\frac {1539545}{
14155776}} \big) {y}^{7}-{\frac {25725}{524288}}\,{y}^{9}
$&$0
$&$\big( {\frac {1485}{8192}}\,{\lambda}^{3}+{\frac {439675}{
1769472}}\,{\lambda}^{2}-{\frac {246977}{28311552}}\,\lambda-{\frac {
513938083}{10192158720}} \big){y}^{3} + \big( {\frac {17435}{442368}}\,{
\lambda}^{2}+{\frac {190295}{14155776}}\,\lambda-{\frac {117973177}{
5096079360}} \big) {y}^{5}+ \big( {\frac {36449}{393216}}\,\lambda+
{\frac {6114493}{94371840}} \big) {y}^{7}-{\frac {8085}{262144}}\,{y
}^{9}
$&$0
$&$\big( {\frac {12385}{1769472}}\,{\lambda}^{2}+{\frac {35497}{
28311552}}\,\lambda-{\frac {181782779}{20384317440}} \big){y}^{5} + \big( 
{\frac {5425}{147456}}\,\lambda+{\frac {2961899}{141557760}} \big) {
y}^{7}-{\frac {3003}{262144}}\,{y}^{9}
$&$0
$&$\big( {\frac {32417}{3538944}}\,\lambda+{\frac {375383}{
169869312}} \big){y}^{7} -{\frac {2145}{1048576}}\,{y}^{9}
$\\\end{tabular}

%% file: alpha.tex
\begin{tabular}{c|*{ 2}{>{\footnotesize\PBS\raggedright}p{0.3cm}}
                  *{ 1}{>{\footnotesize\PBS\raggedright}p{1.2cm}}
                  *{ 1}{>{\footnotesize\PBS\raggedright}p{1.7cm}}
                  *{ 1}{>{\footnotesize\PBS\raggedright}p{2.0cm}}
                  *{ 1}{>{\footnotesize\PBS\raggedright}p{2.6cm}}
                  *{ 1}{>{\footnotesize\PBS\raggedright}p{3.0cm}}
                  *{ 1}{>{\footnotesize\PBS\raggedright}p{3.5cm}}
                  *{ 1}{>{\footnotesize\PBS\raggedright}p{4.5cm}}}
{\Large $ _l\!\!\diagdown\!\!^i$} & 0 & 1 & 2 & 3 & 4 & 5 & 6 & 7 & 8  \\  \hline 
1 & $1
$&$0
$&$0
$&$0
$&${\frac {25}{128}}\,{\lambda}^{2}+{\frac {175}{768}}\,\lambda+{\frac {
1225}{18432}}
$&$0
$&${\frac {25}{64}}\,{\lambda}^{3}+{\frac {68075}{98304}}\,{\lambda}^{2}+
{\frac {410275}{1179648}}\,\lambda+{\frac {2568475}{56623104}}
$&$0
$&${\frac {134925}{131072}}\,{\lambda}^{4}+{\frac {239525}{98304}}\,{
\lambda}^{3}+{\frac {316929175}{169869312}}\,{\lambda}^{2}+{\frac {
1001961695}{2038431744}}\,\lambda+{\frac {2116102111}{97844723712}}
$\\2 & ---&$-\frac{1}{8}
$&$0
$&$-{\frac {25}{32}}\,\lambda-{\frac {175}{384}}
$&$0
$&$-{\frac {475}{512}}\,{\lambda}^{2}-{\frac {555}{512}}\,\lambda-{\frac 
{965}{4608}}
$&$0
$&$-{\frac {10025}{4096}}\,{\lambda}^{3}-{\frac {1698785}{393216}}\,{
\lambda}^{2}-{\frac {28823195}{14155776}}\,\lambda-{\frac {123632507}{
679477248}}
$&$0
$\\3 & ---&---&${\frac {5}{16}}\,\lambda+{\frac {17}{96}}
$&$0
$&${\frac {5}{16}}\,{\lambda}^{2}+{\frac {1315}{3072}}\,\lambda+{\frac {
9235}{73728}}
$&$0
$&${\frac {1765}{2048}}\,{\lambda}^{3}+{\frac {1985}{1024}}\,{\lambda}^{2
}+{\frac {4356505}{3538944}}\,\lambda+{\frac {18119345}{84934656}}
$&$0
$&${\frac {2195}{1024}}\,{\lambda}^{4}+{\frac {3052455}{524288}}\,{
\lambda}^{3}+{\frac {888829145}{169869312}}\,{\lambda}^{2}+{\frac {
515947105}{301989888}}\,\lambda+{\frac {13524934807}{97844723712}}
$\\4 & ---&---&---&$-{\frac {5}{256}}\,\lambda-{\frac {107}{6144}}
$&$0
$&$-{\frac {475}{1536}}\,{\lambda}^{2}-{\frac {323095}{884736}}\,\lambda-
{\frac {2199527}{21233664}}
$&$0
$&$-{\frac {15815}{24576}}\,{\lambda}^{3}-{\frac {6138925}{5308416}}\,{
\lambda}^{2}-{\frac {71128549}{113246208}}\,\lambda-{\frac {
12275264429}{122305904640}}
$&$0
$\\5 & ---&---&---&---&${\frac {25}{768}}\,{\lambda}^{2}+{\frac {515}{13824}}\,\lambda+{\frac 
{13777}{1327104}}
$&$0
$&${\frac {25}{384}}\,{\lambda}^{3}+{\frac {163145}{1327104}}\,{\lambda}^
{2}+{\frac {62041}{786432}}\,\lambda+{\frac {133449991}{7644119040}}
$&$0
$&${\frac {20225}{98304}}\,{\lambda}^{4}+{\frac {81963095}{127401984}}\,{
\lambda}^{3}+{\frac {260962627}{382205952}}\,{\lambda}^{2}+{\frac {
10908318845}{36691771392}}\,\lambda+{\frac {1946363428441}{
44030125670400}}
$\\6 & ---&---&---&---&---&$-{\frac {25}{24576}}\,{\lambda}^{2}-{\frac {3749}{1769472}}\,\lambda-{
\frac {384013}{424673280}}
$&$0
$&$-{\frac {425}{12288}}\,{\lambda}^{3}-{\frac {5290135}{84934656}}\,{
\lambda}^{2}-{\frac {82443283}{2264924160}}\,\lambda-{\frac {
6684148507}{978447237120}}
$&$0
$\\7 & ---&---&---&---&---&---&${\frac {125}{73728}}\,{\lambda}^{3}+{\frac {20765}{7077888}}\,{\lambda
}^{2}+{\frac {77707}{47185920}}\,\lambda+{\frac {6177187}{20384317440}
}
$&$0
$&${\frac {125}{24576}}\,{\lambda}^{4}+{\frac {515945}{42467328}}\,{
\lambda}^{3}+{\frac {1166591837}{101921587200}}\,{\lambda}^{2}+{\frac 
{62067170651}{12230590464000}}\,\lambda+{\frac {3643667826661}{
4109478395904000}}
$\\8 & ---&---&---&---&---&---&---&$-{\frac {25}{1179648}}\,{\lambda}^{3}-{\frac {595}{6291456}}\,{\lambda
}^{2}-{\frac {619081}{6794772480}}\,\lambda-{\frac {28211731}{
1141521776640}}
$&$0
$\\9 & ---&---&---&---&---&---&---&---&${\frac {125}{2359296}}\,{\lambda}^{4}+{\frac {41825}{339738624}}\,{
\lambda}^{3}+{\frac {1276387}{12230590464}}\,{\lambda}^{2}+{\frac {
692812541}{17978967982080}}\,\lambda+{\frac {18307106611}{
3451961852559360}}
$\\\end{tabular}